\documentclass[final,fleqn,10pt]{siamltex}
\usepackage{color}
\usepackage{graphicx}
\usepackage{caption}
\usepackage{float}
\usepackage[lofdepth,lotdepth]{subfig}

\usepackage{sidecap}
\usepackage{url}

\usepackage[final]{listings}
\newcommand{\emcolor}{blue}
\usepackage{amssymb}
\usepackage{amsmath}
\newcommand{\bq}{\begin{equation}}
\newcommand{\eq}{\end{equation}}

\newcommand{\flops}{\mbox{flops}}

\newcommand{\GBS}{\mbox{GB/s}}

\newcommand{\LUPS}{\mbox{LUP/s}}

\newcommand{\GLUPS}{\mbox{GLUP/s}}
\newcommand{\GHZ}{\mbox{GHz}}

\newcommand{\LUP}{\mbox{LUP}}
\newcommand{\LUPs}{\mbox{LUPs}}

\newcommand{\bytes}{\mbox{bytes}}
\newcommand{\byte}{\mbox{byte}}
\newcommand{\bit}{\mbox{bit}}

\newcommand{\GB}{\mbox{GB}}

\newcommand{\GiB}{\mbox{GiB}}
\newcommand{\KiB}{\mbox{KiB}}
\newcommand{\MiB}{\mbox{MiB}}

\newcommand{\eos}{~.}
\newcommand{\cma}{~,}
\newcommand{\lperfctr}{\texttt{likwid-perfctr}}

\newcommand{\construction}[1]{}

\newcommand{\ttomega}{\mbox{$\mathtt{\Omega}$}}

\makeatletter
\let\@fnsymbol\@arabic
\makeatother

\begin{document}

\title{Multicore-optimized wavefront diamond blocking for optimizing stencil updates}
\author{T. Malas\thanks{Extreme Computing Research Center (ECRC), King Abdullah University of Science and Technology (KAUST), Saudi Arabia} \and G. Hager\thanks{Erlangen Regional Computing Center (RRZE), Friedrich-Alexander University of Erlangen-Nuremberg, Germany} \and H. Ltaief\footnotemark[1] \and H. Stengel\footnotemark[2] \and G. Wellein\footnotemark[2] \and D. Keyes\footnotemark[1]}
\date{\today}
\maketitle
\begin{abstract}
The importance of stencil-based algorithms in computational science
has focused attention on optimized parallel implementations for 
multilevel cache-based processors. Temporal blocking
schemes leverage the large bandwidth and low latency of caches to 
accelerate stencil updates and approach theoretical peak 
performance. A key ingredient is the reduction of data traffic 
across slow data paths, especially the main memory interface.  In 
this work we combine the ideas of multi-core wavefront temporal 
blocking and diamond tiling to arrive at stencil update 
schemes that show large reductions in memory pressure compared to 
existing approaches. 
The resulting schemes show performance advantages in bandwidth-starved situations, which are exacerbated by the high bytes per lattice update case of variable coefficients.
Our thread groups concept provides a controllable trade-off between concurrency and memory usage, shifting the pressure between the memory interface and the CPU.
We present performance results on a contemporary Intel processor.
\end{abstract}

\section{Introduction}\label{sec:intro}

\subsection{Stencil codes}
The evaluation of stencil operators on Cartesian lattices is a classic kernel in computational science and engineering, arising from systems that are “born” discrete and from discretizations of PDEs, both explicit and implicit.  In the implicit case the iteration index is analogous to explicit time and stencil evaluation becomes a case of sparse matrix-vector multiplication with special structure.  Lattice values are updated from neighbors at a previous level with concurrency that scales linearly with the number of degrees of freedom.  However, low flop-per-byte ratios put a premium on locality. Regular access patterns allow high spatial locality, in the sense of packing cache blocks.  The modest temporal locality within a single iteration level from reuse of a value within several adjacent stencils can be enhanced across iteration levels. 

A major demarcation exists between stencils whose coefficients are constant in space and time and those that vary, since variable coefficients can shift the dominant workingset from the lattice values being updated to the coefficients, themselves.  In the PDE context, coefficient variability can arise from constitutive parameters (conductivities, elastic moduli, etc.) that depend upon space or time intrinsically or through dependences on the evolving field values, themselves, the typical nonlinear case.  Some models can be scaled so that the variability affects only the diagonal term of the stencil, which is then an important case to which to specialize. Whether to compute coefficients on-the-fly is a decision that affects the hardware resource balance, since it both releases all of the memory bandwidth to the lattice values and increases the flop intensity of the typical lattice update. 

Locality is also affected in a major way by the truncation order of the discretization, which manifests itself in the stencil size, each lattice value appearing in more stencil evaluations.  Trends in PDE modeling towards high-order discretizations and high fidelity physics shift interest beyond the low-order constant-coefficient models to which most computer science optimizations have been directed to date.  For instance, seismic models used in oil exploration are typically eighth-order and variable coefficients are the norm.  Traditional means of obtaining high truncation order by expanding the spatial extent of the stencil tax distributed memory parallelization of stencil evaluation by expanding the halo region.  More spatially compact means of obtaining high order are therefore of interest.

The aforementioned issues put once-humble stencil evaluation in the cross-hairs of co-design and motivate our examination of several shared-memory (multi-core) and distributed-memory (message-passing) optimizations of a variety of stencils (see Figure~\ref{fig:naive_stencil_codes}) on state-of-the-art hardware.  We are especially concerned with the degradation of memory bandwidth per core forecast at tomorrow's extreme scale.  We combine classical and novel techniques and test them on a range of star-like stencils accommodating up to eighth-order, constant and variable coefficient (without on-the-fly recomputation), noting their salutary effects on memory pressure, power consumption, and obtainable performance, and noting the transition of hardware bottlenecks.

This paper merely scratches the surface of co-design for the fundamental kernel of Cartesian lattice updates.  Pipelined or s-step Krylov solvers, time-parallelism, and high-order temporal discretizations that are obtained by using the governing PDE to estimate high time derivatives with high space derivatives are all potentially stencil-expanding (and halo-expanding) decisions made at an upstream algorithmic stage.  Their downstream consequences on stencil update performance and hardware balance can be examined using the analyses and software tools introduced herein, but it remains to close the loop with analyses and tools that allow how to design the best discrete schemes in the first place.

\subsection{Contribution}

This work makes the following contributions:
We combine two previously known concepts for optimizing stencil 
  computations, diamond tiling and multi-core aware temporal wavefront 
  blocking, and show our method to be more efficient in terms of 
  flexibility, cache requirements, and main memory pressure than 
  either concept by itself on a modern Intel 10-core CPU.
Our method shows best performance for stencil types with low
  computational intensity, such as long-range stencils with variable
  coefficients. The reduced memory bandwidth pressure has a positive
  side effect of significant energy savings in memory.
We also show that MPI parallelization of the method is straightforward
  and enables natural overlapping of computation and communication.

The rest of this paper is organized as follows: After an overview of
related work in Sect.~\ref{sec:related} we introduce the test system,
define some basic terminology, and study a variety of stencil update
schemes together with baseline performance data without temporal
blocking but otherwise minimal code balance in Sect.~\ref{sec:RLmodel}. In
Sect.~\ref{sec:tblocking} we briefly review multi-core wavefront
temporal blocking and diamond tiling before elaborating on the
combination of the two concepts in Section~\ref{approach}. In 
Sect.~\ref{sec:results} we present performance results, 
and Sect.~\ref{sec:conc} provides a summary and an outlook
to possible future work.

\section{Related work}\label{sec:related}
The importance of stencil computations and the inefficient performance of 
their na\"{i}ve implementations on modern processors motivates 
researchers to study them extensively.  The optimizations required to 
achieve the desired performance depend on the properties of the stencil 
operator and the capabilities of different resources in the processor.  
This case is made by Datta~\cite{Datta:EECS-2009-177}, where the 
performance of several combinations of optimization techniques, 
processors, and stencil operators is reported.

The high \bytes\ per lattice site update (\LUP) requirement of many stencil computations and the increasing performance gap between the arithmetic operations and the data transfer are the major concerns in achieving high performance.
Spatial and temporal blocking improve the performance by increasing the data reuse in the cache memory of modern processors.

Spatial blocking is an established technique that changes the grid traversal order to maximise the data reuse in the desired memory level \cite{datta09, dursun2012hierarchical}.
Temporal blocking allows more data reuse in the cache memory by reordering grid traversal across space iterations, where blocks of grid points are accessed multiple times before completing the traversal of a single spatial grid level.

Temporal blocking techniques require careful handling of data dependencies acoss space iterations to ensure correctness.  
Several tiling techniques are proposed in the literature including: parallelepiped, split, overlapped, diamond, and hexagonal.
These block shapes optimize for data locality, concurrency, or both.  
Reviews of these techniques can be found at Orozco \textit{et al.}~\cite{orozco2011locality} and Zhou~\cite{Zhou2013}.  
We believe that diamond tiling is promising for efficiently providing both concurrency and data locality over the problems and computer architectures of our interest.
Its attractiveness in recent years is evident in: \cite{orozco2011locality}, \cite{Zhou2013}, Strzodka \textit{et al.}~\cite{Strzodka:2011:CAT}, Bandishti~\textit{et al.} \cite{Bandishti6468470}, and Grosser \textit{et al.}~\cite{grosser2014hybrid}, where a GPU implementation of hexagonaltiling is proposed, then a study of hexagonal and diamond tiling is performed~\cite{grosser2014ppl}.

The wavefront technique, which is introduced by Lamport~\cite{Lamport:1974} (using the name ``hyperplane''), performs temporal blocking at adjacent grid points.
This technique has been combined with other tiling approaches using single-thread wavefront temporal blocking as in~\cite{Strzodka:2011:CAT}, Wonnacott  \textit{et al.}~\cite{Wonnacott845979}, and Nguyen \textit{et al.}~\cite{nguyen18993}, and using multi-threaded wavefront temporal blocking, as in Wellein \textit{et al.}~\cite{wellein5254211}.

Cache optimization techniques can be classified into cache-oblivious and cache-aware techniques. Cache-oblivious techniques ~\cite{Frigo:2005:COS:1088149.1088197, Tang2011, Strzodka:2010:COP:1810085.1810096} do not need prior knowledge or tuning to find optimal cache block size to achieve high performance stencil computations.
On the other hand, cache-aware techniques utilize auto-tuning as in~\cite{Datta:EECS-2009-177}, which performs parameter search over the optimization techniques to achieve best performance.
Another cache-aware algorithm is introduced in~\cite{Strzodka:2011:CAT}, where cache block size calculations are used to set the cache block size that achieves best performance.

Several frameworks have been developed to produce optimized stencil codes.
PLUTO~\cite{Bondhugula2008} is a source-to-source transformation tool that uses polyhedral model, CATS~\cite{Strzodka:2011:CAT} is a library, Pochoir~\cite{Tang2011} uses cache-oblivious algorithms  in Domain Specific Languages (DSL), PATUS~\cite{Christen6012879} uses auto-tuning with a DSL, and Henretty \textit{et al.}~\cite{henretty2013stencil} develop a DSL that uses split-tiling. 
Unat \textit{et al.}~\cite{unat2011mint} introduced Mint, a programming model that produces highly optimized GPU code from a user's annotated traditional C code.
Physis, a DSL that generates optimized GPU codes with the necessary MPI calls for heterogeneous GPU clusters, was proposed by Maruyama \textit{et al.}~\cite{maruyama2011physis}.
A recent review paper of stencil optimization tools that use polyhedral model has been prepared by Wonnacott~\cite{Wonnacott2013}.

\section{Detailed analysis of various stencil operators}\label{sec:RLmodel}

\subsection{Definitions and terminology}

In iterative stencil computations, each point in a multi-dimensional
spatial grid ($\Omega$) is updated using weighted contributions from its
neighbor points, defined by the stencil operator.  The stencil
operator specifies the relative coordinates of the contributing points
and their weights.  The weights can be constant or variable in space and/or time with some
or no symmetry to be exploited around the updated point.  The grid update operation
over the complete spatial domain (one ``sweep'') is repeated \verb.T. times (time
steps or iterations).  We consider
three-dimensional grids in this work ($\Omega:=\{1,\dots
,N_x\}\times\{1,\dots ,N_y\}\times \{1,\dots ,N_z\}$), where
$x$ is the leading dimension, followed by the $y$ and
$z$ dimensions.  A grid location \verb.(x1,y1,z1). at time step
\verb.t. can be expressed as $\Omega^t_{z_1,y_1,x_1}$.  We also follow
the common practice regarding the mapping of the spatial and temporal
domain to a data structure in the C language: $\Omega^t_{z,y,x}$ is
stored as \ttomega\verb.[t][z][y][x]., where the spatial indices
\verb.x., \verb.y., and \verb.z. can assume integer values in the
range $0\ldots N_i-1$ so that the innermost loop accesses memory with a
stride of one.  The time index can
only assume the values $0$ and $1$, so the data from time step $t-1$ 
is overwritten by the new data for $t+1$ in each grid update.  
PDE discretizations higher than first order in temporal truncation error can often be accommodated by employing the original underlying PDE to replace higher-order-in-time derivatives in the truncation error with their spatial equivalent. This expands the spatial stencil while keeping the number of domain copies at minimum.
Figure~\ref{fig:naive_stencil_codes} shows several
examples of stencil computation kernels that use a ``Jacobi-like'' update scheme
where the stencil array (i.e., the data structure with read access to 
neighboring grid points) is not written to during the same sweep.
Another possible variant is ``Gauss-Seidel'' update scheme, where the stencil array is adapted during the same sweep. The latter is relevant in practice but beyond the scope of this paper. All stencils considered here are ``star stencils'' of 
various spatial differential or truncation orders. ``Box 
stencils'', ``diamond stencils'', and multicomponent stencils, in which multiple discrete fields interact on the same (or interlaced, staggered) lattices, are also important in practice, and can be handled with similar techniques.

\begin{figure*}[tbp]
\centering
\subfloat[$1^{st}$ order in time 3-point constant-coefficient stencil in one dimension, with symmetry.]{
\def\arraystretch{1.2}
\begin{tabular}{|l|}
\hline
$for\ t = 0\ to\ T-1$\\
\ \ $for\ x = 1\ to\ N_x$\\
\ \ \ \ $\Omega^{t+1}_{x} = w_1 * \Omega^{t}_{x}$\\
\ \ \ \ \ \ $+ w_2 * (\Omega^{t}_{x-1} + \Omega^{t}_{x+1})$\ \ \ \ \\
\\
\\
\\
\\

\hline
\end{tabular}
\label{fig:3_pt_stencil_1d}
}
\quad
\subfloat[$1^{st}$ order in time 7-point constant-coefficient isotropic stencil in three dimensions, with symmetry.]{
\def\arraystretch{1.2}
\begin{tabular}{|l|}
\hline
$for\ t = 0\ to\ T-1$\\
\ \ $for\ z = 1\ to\ N_z$\\
\ \ \ \ $for\ y = 1\ to\ N_y$\\
\ \ \ \ \ \ $for\ x = 1\ to\ N_x$\\
\ \ \ \ \ \ \ \ $\Omega^{t+1}_{z,y,x} = w_1 * \Omega^{t}_{z,y,x}$\\
\ \ \ \ \ \ \ \ \ \ $+ w_2 *( \Omega^{t}_{z,y,x-1} + \Omega^{t}_{z,y,x+1})$\ \ \ \ \ \ \ \ \ \\
\ \ \ \ \ \ \ \ \ \ $+ w_2 *( \Omega^{t}_{z,y-1,x} + \Omega^{t}_{z,y+1,x})$\\
\ \ \ \ \ \ \ \ \ \ $+ w_2 *( \Omega^{t}_{z-1,y,x} + \Omega^{t}_{z+1,y,x})$\\
\hline

\end{tabular}
\label{fig:7_pt_stencil_3d}
}

~

\subfloat[$1^{st}$ order in time 7-point variable-coefficient stencil in three dimension, with no coefficient symmetry.]{
\def\arraystretch{1.2}
\begin{tabular}{|l|}
\hline
$for\ t = 0\ to\ T-1$\\
\ \ $for\ z = 1\ to\ N_z$\\
\ \ \ \ $for\ y = 1\ to\ N_y$\\
\ \ \ \ \ \ $for\ x = 1\ to\ N_x$\\
\ \ \ \ \ \ \ \ $\Omega^{t+1}_{z,y,x} =$\\
\ \ \ \ \ \ \ \ \ \ \ \ $W_{0,z,y,x} * \Omega^{t}_{z,y,x}$\\
\ \ \ \ \ \ \ \ \ \ $+ W_{1,z,y,x} *\Omega^{t}_{z,y,x-1}$\\
\ \ \ \ \ \ \ \ \ \ $+ W_{2,z,y,x} *\Omega^{t}_{z,y,x+1}$\\
\ \ \ \ \ \ \ \ \ \ $+ W_{3,z,y,x} *\Omega^{t}_{z,y-1,x}$\\
\ \ \ \ \ \ \ \ \ \ $+ W_{4,z,y,x} *\Omega^{t}_{z,y+1,x}$\\
\ \ \ \ \ \ \ \ \ \ $+ W_{5,z,y,x} *\Omega^{t}_{z-1,y,x}$\\
\ \ \ \ \ \ \ \ \ \ $+ W_{6,z,y,x} *\Omega^{t}_{z+1,y,x}$\\
\\
\\
\\
\\
\\
\hline
\end{tabular}
\label{fig:7_pt_var_stencil_3d}
}
\quad
\subfloat[$1^{st}$ order in time 25-point variable-coefficient anisotropic stecil in three dimensions, with symmetry across each axis.]{
\def\arraystretch{1.2}
\begin{tabular}{|l|}
\hline
$for\ t = 0\ to\ T-1$\\
\ \ $for\ z = 1\ to\ N_z$\\
\ \ \ \ $for\ y = 1\ to\ N_y$\\
\ \ \ \ \ \ $for\ x = 1\ to\ N_x$\\
\ \ \ \ \ \ \ \ $\Omega^{t+1}_{z,y,x} = W_{0,z,y,x} * \Omega^{t}_{z,y,x}$\\
\ \ \ \ \ \ \ \ \ \ $+ W_{1,z,y,x} *( \Omega^{t}_{z,y,x-1} + \Omega^{t}_{z,y,x+1})$\\
\ \ \ \ \ \ \ \ \ \ $+ W_{2,z,y,x} *( \Omega^{t}_{z,y-1,x} + \Omega^{t}_{z,y+1,x})$\\
\ \ \ \ \ \ \ \ \ \ $+ W_{3,z,y,x} *( \Omega^{t}_{z-1,y,x} + \Omega^{t}_{z+1,y,x})$\\

\ \ \ \ \ \ \ \ \ \ $+ W_{4,z,y,x} *( \Omega^{t}_{z,y,x-2} + \Omega^{t}_{z,y,x+2})$\\
\ \ \ \ \ \ \ \ \ \ $+ W_{5,z,y,x} *( \Omega^{t}_{z,y-2,x} + \Omega^{t}_{z,y+2,x})$\\
\ \ \ \ \ \ \ \ \ \ $+ W_{6,z,y,x} *( \Omega^{t}_{z-2,y,x} + \Omega^{t}_{z+2,y,x})$\\

\ \ \ \ \ \ \ \ \ \ $+ W_{7,z,y,x} *( \Omega^{t}_{z,y,x-3} + \Omega^{t}_{z,y,x+3})$\\
\ \ \ \ \ \ \ \ \ \ $+ W_{8,z,y,x} *( \Omega^{t}_{z,y-3,x} + \Omega^{t}_{z,y+3,x})$\\
\ \ \ \ \ \ \ \ \ \ $+ W_{9,z,y,x} *( \Omega^{t}_{z-3,y,x} + \Omega^{t}_{z+3,y,x})$\\

\ \ \ \ \ \ \ \ \ \ $+ W_{10,z,y,x} *( \Omega^{t}_{z,y,x-4} + \Omega^{t}_{z,y,x+4})$\\
\ \ \ \ \ \ \ \ \ \ $+ W_{11,z,y,x} *( \Omega^{t}_{z,y-4,x} + \Omega^{t}_{z,y+4,x})$\\
\ \ \ \ \ \ \ \ \ \ $+ W_{12,z,y,x} *( \Omega^{t}_{z-4,y,x} + \Omega^{t}_{z+4,y,x})$\\
\hline

\end{tabular}
\label{fig:25_pt_var_stencil_3d}
}

~

\subfloat[$2^{nd}$ order in time 25-point constant-coefficient isotropic stencil in three dimensions, with symmetry across each axis.]{
\def\arraystretch{1.2}
\begin{tabular}{|l|}
\hline
$for\ t = 0\ to\ T-1$\\
\ \ $for\ z = 1\ to\ N_z$\\
\ \ \ \ $for\ y = 1\ to\ N_y$\\
\ \ \ \ \ \ $for\ x = 1\ to\ N_x$\\
\ \ \ \ \ \ \ \ $\Omega^{t+1}_{z,y,x} = 2*\Omega^{t}_{z,y,x} - \Omega^{t-1}_{z,y,x} + \alpha_{z,y,x}*[w_0 * \Omega^{t}_{z,y,x}$\\
\ \ \ \ \ \ \ \ \ \ $+ w_1 *(( \Omega^{t}_{z,y,x-1} + \Omega^{t}_{z,y,x+1})$
                    $+       ( \Omega^{t}_{z,y-1,x} + \Omega^{t}_{z,y+1,x})$
                    $+       ( \Omega^{t}_{z-1,y,x} + \Omega^{t}_{z+1,y,x}))$\\
\ \ \ \ \ \ \ \ \ \ $+ w_2 *(( \Omega^{t}_{z,y,x-2} + \Omega^{t}_{z,y,x+2})$
                    $+       ( \Omega^{t}_{z,y-2,x} + \Omega^{t}_{z,y+2,x})$
                    $+       ( \Omega^{t}_{z-2,y,x} + \Omega^{t}_{z+2,y,x}))$\\
\ \ \ \ \ \ \ \ \ \ $+ w_3 *(( \Omega^{t}_{z,y,x-3} + \Omega^{t}_{z,y,x+3})$
                    $+       ( \Omega^{t}_{z,y-3,x} + \Omega^{t}_{z,y+3,x})$
                    $+       ( \Omega^{t}_{z-3,y,x} + \Omega^{t}_{z+3,y,x}))$\\
\ \ \ \ \ \ \ \ \ \ $+ w_4 *(( \Omega^{t}_{z,y,x-4} + \Omega^{t}_{z,y,x+4})$
                    $+       ( \Omega^{t}_{z,y-4,x} + \Omega^{t}_{z,y+4,x})$
                    $+       ( \Omega^{t}_{z-4,y,x} + \Omega^{t}_{z+4,y,x}))]$\\
\hline

\end{tabular}
\label{fig:25_pt_const_stencil_3d}
}

\caption{Examples of iterative stencil computations, starting with
  initial domain $\Omega^0$.  The stencil operators in b, c, d, and e
  are analyzed in this paper. $w$ represents scalar coefficients and $W$ represents domain-sized coefficients.}
\label{fig:naive_stencil_codes}
\end{figure*}

\subsection{Test systems}

All benchmark tests were performed on a cluster of dual-socket Intel
Ivy Bridge (Xeon E5-2660v2) nodes with a nominal clock speed of
2.2\,\GHZ\ and ten cores per chip.  The ``Turbo Mode'' feature was
disabled. Each CPU has a 25\,\MiB\ L3 cache which is shared among all
cores, and core-private L2 and L1 caches of 256\,\KiB\ and 32\,\KiB,
respectively.  All data paths between the cache levels are
half-duplex, 256-\bit\ wide buses, so the transfer of one
64-\byte\ cache line between adjacent caches takes two CPU cycles. The
core architecture supports all Intel Single Instruction Multiple Data
(SIMD) instruction sets up to AVX (Advanced Vector Extensions).
With AVX, one core is able to sustain one full-width (32\,\byte) 
load and one half-width (16\,\byte) store per cycle. In addition,
one AVX multiply and one AVX add instruction can be executed per cycle.
Since one AVX register can hold either four double precision (DP) or
eight single precision (SP) operands, the peak performance of one
core is eight \flops\ per cycle in DP or sixteen \flops\ per cycle in 
SP. 

Each node is equipped with 64\,\GB\ of DDR3-1600 RAM per socket and
has a maximum attainable memory bandwidth of $b_\mathrm S\approx 40\,\GBS$ per
socket (as measured with the STREAM COPY \cite{stream} \cite{McCalpin1995} benchmark).
The nodes are connected by a full non-blocking, fat-tree QDR
InfiniBand network.

For compiling and linking, the Intel C compiler in version 13.1.3 was
used together with the Intel MPI library 4.1.3. Hardware performance
counter measurements were done with \texttt{likwid-perfctr}
from the LIKWID multicore tools collection \cite{likwid}.

\subsection{Performance prediction and measurements for pure spatial blocking}

In this section we analyze two basic stencil update schemes
for their requirements on the
hardware. It is not our goal to provide a comprehensive coverage of
all possible variations in stencil algorithms; we rather wish to
motivate that our selection includes relevant corner cases. 
We use (\LUPS) as a
basic performance metric, since it does not contain any uncertainty
as to how many \flops\ are actually done during one stencil update.
Specific implementations have a fixed ratio of \LUPs\ to \flops\ and 
other relevant hardware events (such as bytes transferred, instructions
executed, etc.), which are discussed as required. Unless otherwise 
noted, the working set does not fit into any CPU cache.

In the following we describe in detail two ``corner
cases'' of stencil update schemes: a three-dimensional seven-point stencil with
constant coefficients (Jacobi-type smoother, see
Fig.~\ref{fig:7_pt_stencil_3d}) and a three-dimensional 25-point stencil with
constant coefficients (see Fig.~\ref{fig:25_pt_const_stencil_3d}). 
These examples were picked
because they are simple to model for memory-bound situations but only
the first is a good candidate for temporal blocking, as will be shown in the following two subsections. Later we will
show the effectiveness of temporal blocking for the Jacobi smoother, the
variable-coefficient 7-point stencil shown in
Fig.~\ref{fig:7_pt_var_stencil_3d}, and the variable-coefficient 25-point
stencil shown in Fig.~\ref{fig:25_pt_var_stencil_3d}.

\subsubsection{Three-dimensional seven-point stencil with constant coefficients}

The standard two-grid three-dimensional ``Jacobi'' update scheme in
Fig.~\ref{fig:7_pt_stencil_3d} is probably the best analyzed stencil
algorithm to date. From a data flow perspective the spatial loop nest
reads one array (\ttomega\verb.[0][][][].) and updates another
(\ttomega\verb.[1][][][].). In double precision, the minimum code balance
is thus $B_\mathrm C=24\,\bytes/\LUP$: eight \bytes\ for loading one new element of the
previous time step data, eight bytes for the write-allocate transfer on
the new time step, and eight bytes for evicting the updated data back
to memory. The write-allocate transfer may be avoided by the use of
``non-temporal stores,'' which bypass the memory hierarchy, thereby
reducing the code balance to $16\,\bytes/\LUP$. Since this optimization 
is of minor importance for the practically relevant cases shown below,
we will not discuss it here any further.

Depending on the grid size and the cache size, spatial blocking may
be required to achieve the minimum code balance of
$24\,\bytes/\LUP$. If three successive ``layers'' of size $N_x\times
N_y$ grid points fit into a cache, the only load operation within a
lattice site update that causes a cache miss goes to
\ttomega\verb.[0][z+1][y][x]., and all other loads can be satisfied from
the cache. If $C$ is the cache size, we assume (as a rule of thumb) that
only about $C/2$ is available for the previous 
time step data, and the layer condition for double precision is
\bq\label{eq:lcond_7pt_const}
3\times N_x\times N_y\times 8\,\bytes < \frac C{2n_\mathrm{threads}}\eos
\eq
This assumes that OpenMP parallelization is done along the $z$ axis
with static scheduling.
If this condition is violated, at least the loads to  \ttomega\verb.[0][z+1][y][x].,
\ttomega\verb.[0][z-1][y][x]., and \ttomega\verb.[0][z][y+1][x].
will cause cache misses, which leads to a code balance of 
$40\,\bytes/\LUP$. If the cache is too small to even hold
three successive rows of the grid, the only loads that come
from the cache will be to  \ttomega\verb.[0][z][y][x-1].  and
to \ttomega\verb.[0][z][y][x].. The code balance for this case
is $56\,\bytes/\LUP$.

The layer condition (\ref{eq:lcond_7pt_const}) is independent of
$N_z$. Hence, it is sufficient to introduce spatial blocking in the
$x$ and/or $y$ dimensions in order to arrive at the minimum code
balance.  In practice one should try to keep the inner ($x$) block
size larger than about one OS page in order to avoid frequent TLB
misses and excess data traffic due to hardware
prefetching~\cite{hpc4se}. 
Additionally, we use ``static,1'' OpenMP scheduling, which relaxes the layer
condition to
\bq\label{eq:lcond_7pt_const_static1}
\left(n_\mathrm{threads}+2\right)\times N_x\times N_y\times 8\,\bytes < \frac C{2}\cma
\eq
since each thread shares both neighboring layers of its current $z$
layer with its neighboring threads (except the first and the last
thread, which only share one layer with their respective neighbor).

Note that the layer condition can be
satisfied for any cache in the hierarchy if the block sizes are chosen
appropriately; for memory-bound implementations one usually tries to
establish it for the outer-level cache (OLC) to ameliorate the impact
of the memory bandwidth bottleneck.  In case of temporal blocking,
however, the bottleneck may not be main memory and the smaller caches
need to be taken into account. Since the overhead at block boundaries
becomes significant at small block sizes, the optimum code balance is
a goal that is all but impossible to achieve in this case.
 
Figure \ref{fig:7_pt_const_naive_large} shows the performance of the
seven-point stencil algorithm in double precision on one Ivy Bridge
chip with up to ten cores for a grid of $960^3$ points
(circles) and the memory bandwidth as measured by
\texttt{likwid-perfctr} (triangles), together with the estimated saturated performance (solid line) and ideal scaling (dashed line). With appropriate spatial blocking
the expected saturated performance as given by the roof\/line model is 
\bq
P_\mathrm{roof} = \frac{b_\mathrm S}{B_\mathrm C} = \frac{40\,\GBS}{24\,\bytes/\LUP} =
1.67\,\GLUPS\eos
\eq
The performance saturates at $6$--$7$ cores, and the available memory
bandwidth is utilized by up to 95\%. Since there is strong saturation,
we expect a strong benefit from temporal blocking.

\subsubsection{Three-dimensional 25-point stencil with constant coefficients}

The considerations about layer conditions as shown above for the
seven-point stencil apply in a similar way for long-range stencils.
In the particular case of the algorithm shown in
Fig.~\ref{fig:25_pt_const_stencil_3d}, one sweep of the grid updates
one array (read/modify/write) and reads two more arrays, one of which
is accessed in a radius-four (semi-bandwidth of four) stencil pattern. The minimum code balance for
double precision is thus $B_\mathrm C=32\,\bytes/\LUP$.

Due to the long-range stencil the layer condition is changed as compared
to the previous case. With ``static,1'' scheduling, each thread can share
the eight neighboring layers \ttomega\verb.[0][z-4][][].\ldots\ttomega\verb.[0][z-1][][].
and \ttomega\verb.[0][z+1][][].\ldots\ttomega\verb.[0][z+4][][]. with its eight
neighboring threads (four in either $z$ direction), but the top and bottom
threads have less sharing. Consequently, the layer condition is
\bq\label{eq:lcond_25pt_const_static1}
\left(n_\mathrm{threads}+8\right)\times N_x\times N_y\times 8\,\bytes < \frac C{2}\eos
\eq
If this condition is fulfilled,  \ttomega\verb.[z+4][y][x].
is the only element from the stencil array that has to come from
main memory.

Figure~\ref{fig:25_pt_const_naive_large} shows the performance and
memory bandwidth for ideal spatial blocking on a ten-core Ivy Bridge
chip. The roof\/line model predicts an upper performance limit
of
\bq
P_\mathrm{roof} = \frac{b_\mathrm S}{B_\mathrm C} = \frac{40\,\GBS}{32\,\bytes/\LUP} =
1.25\,\GLUPS\eos
\eq
In contrast to the Jacobi-type stencil there is no clear saturation.
The data transfers within the cache hierarchy and the execution of the
loop code with data from the L1 cache take so much time that there is
no sufficient pressure on the memory interface to saturate the
bandwidth even with ten cores, which makes this stencil a bad candidate
for temporal blocking unless there is an opportunity to save significant
time with more efficient in-core execution. A thorough analysis of this effect
would exceed the scope of this work and will be published 
elsewhere~\cite{stengel14}.

\subsubsection{Other stencils}

\begin{figure*}[tbp]
    \centering
    \subfloat[7-point constant-coefficient stencil.]{
        \centering
        \includegraphics[width=3.1cm]{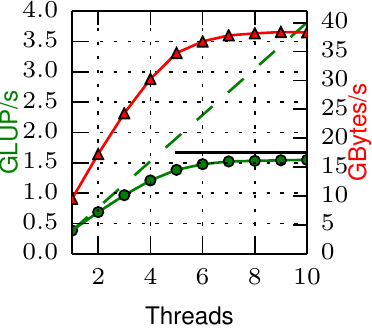}
    \label{fig:7_pt_const_naive_large}
	}
	\enskip
    \subfloat[7-point variable-coefficient stencil.]{
        \centering
        \includegraphics[width=2.75cm]{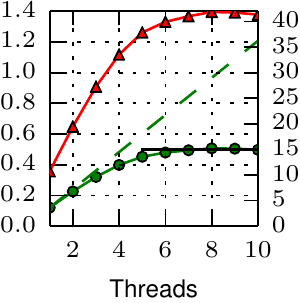}
    \label{fig:7_pt_var_naive_large}
	}
	\enskip
	\subfloat[25-point constant-coefficient stencil.]{
        \centering
        \includegraphics[width=2.75cm]{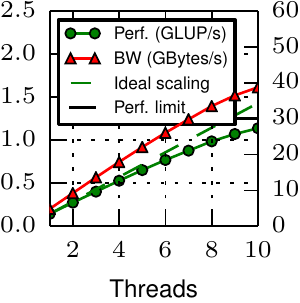}
    \label{fig:25_pt_const_naive_large}
	}
	\enskip
    \subfloat[25-point variable-coefficient stencil.]{
    \centering
        \includegraphics[width=2.75cm]{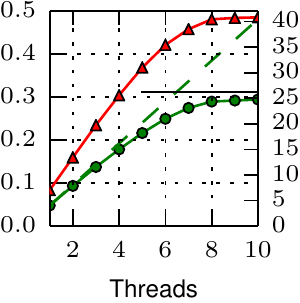}
    \label{fig:25_pt_var_naive_large}
	}
    \caption{Performance scaling across the cores of a chip with
      purely spatial blocking and data sets larger than L3 cache
      for the stencil algorithms shown in Fig.~\ref{fig:naive_stencil_codes}. 
      Problem sizes:
      $960^3$, $680^3$, $960^3$, and $480^3$ for subfigures
      a, b, c, and d, respectively.  STREAM COPY memory bandwidth $b_\mathrm S\approx 40\,\GBS$.} 
\end{figure*}

Figures~\ref{fig:7_pt_var_naive_large} and
\ref{fig:25_pt_var_naive_large} show the saturation characteristics
and maximum performance levels for the seven-point stencil with
variable coefficients (ideal code balance of $80\,\bytes/\LUP$) and
the 25-point stencil with axis-symmetric variable coefficients (ideal
code balance of $128\,\bytes/\LUP$) listed in
Figs.~\ref{fig:7_pt_var_stencil_3d} and
\ref{fig:25_pt_var_stencil_3d}, respectively. Both show strong
saturation close to the performance levels predicted by the roof\/line
model, and are thus viable targets for temporal
blocking optimizations.

\subsection{Upper performance bounds for in-cache execution}

\begin{figure*}[tbp]
    \centering
    \subfloat[7-point constant-coefficient stencil.]{
    \centering
        \includegraphics[width=3.cm]{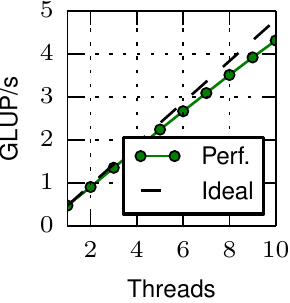}
    \label{fig:7_pt_const_inL3}
	}
	\enskip
    \subfloat[7-point variable-coefficient stencil.]{
    \centering
        \includegraphics[width=2.75cm]{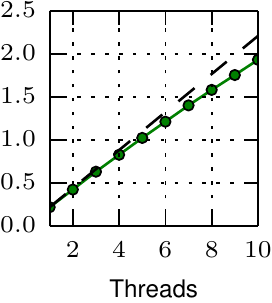}
    \label{fig:7_pt_var_inL3}
	}
	\enskip
	\subfloat[25-point constant-coefficient stencil.]{
    \centering
        \includegraphics[width=2.75cm]{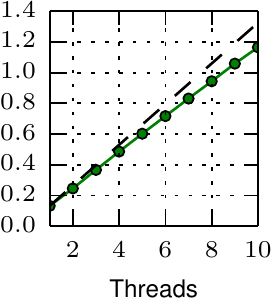}
    \label{fig:25_pt_const_inL3}
	}
	\enskip
    \subfloat[25-point variable-coefficient stencil.]{
    \centering
        \includegraphics[width=2.75cm]{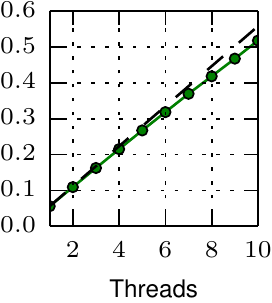}
    \label{fig:25_pt_var_inL3}
	}
    \caption{Performance scaling across the cores of a chip without
      blocking and data sets fitting in the L3 cache
      for the stencil algorithms shown in Fig.~\ref{fig:naive_stencil_codes}. 
      Problem sizes:
      $96\! \times\! 96\! \times\! 96$, $64\! \times\! 64\! \times\! 48$, $128\! \times\! 64\!
      \times\! 64$, and $64\! \times\! 32\! \times\! 32$ for subfigures
      a, b, c, and d, respectively. Smaller grids fit in L3 at variable-coefficient stencils because they require more bytes per grid point to hold the coefficeints values.}
\end{figure*}

To find the expected performance of ideal temporal blocking (i.e.,
when performance has completely decoupled from the memory
bottleneck), we have measured the performance at problems fitting
completely in the last-level cache without temporal blocking.  The
results for the stencils discussed in the previous section are shown
in Figs.~\ref{fig:7_pt_const_inL3}--\ref{fig:25_pt_var_inL3}.  Problem
sizes have been chosen so that work decomposition across threads is
easy (no ``artificial'' load imbalance) and the inner loop length is
not too short.

We see that all stencil algorithms scale very well across the cores,
which is expected since the Ivy Bridge architecture does not have a
hardware bottleneck except the main memory interface. It also shows
that our implementation has no serious issues with OpenMP overhead or
load balancing even with in-cache data sets.

\section{Wavefront and diamond tiling temporal blocking}\label{sec:tblocking}

\subsection{Wavefront temporal blocking}\label{sec:wavefront}
Typical applications have larger grid size than a processor's cache
memory.  If one sweep is completed before the next starts, each time
step involves loading each grid point and writing the result from/to
main memory.  We call this the ``na\"{i}ve'' approach.  As a result,
``na\"{i}ve'' stencil computations are typically memory-bound due to their
low \flops/\byte\ ratio ~\cite{datta09} (see Sect.~\ref{sec:RLmodel}
for a quantitative analysis).  Temporal blocking alleviates the memory
pressure by increasing the in-cache data reuse, i.e.,
several time step updates are performed to a grid point before
evicting the data to main memory.

\begin{figure*}[tbp]
    \centering
    \subfloat[na\"{i}ve: Lattice sites updated in chronological order, one time step at a time. Data dependencies example represented by red arrows.]{
    \centering
        \includegraphics[clip=true, trim = 0.0cm 0.0cm 0.0cm 0.0cm ,width=3.5cm]{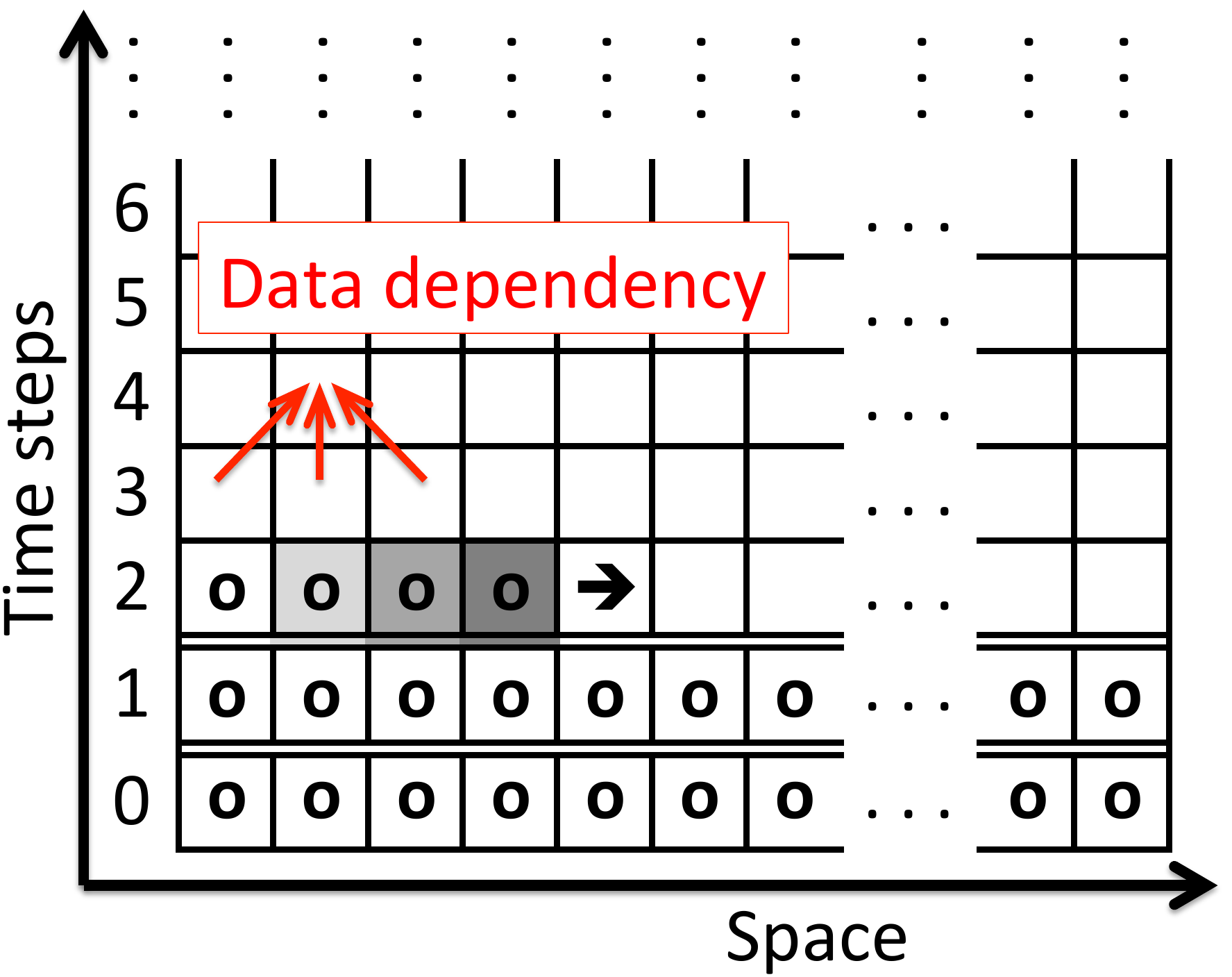}
    \label{fig:naive_3pt_1d}
    }
     \quad
    \subfloat[Single thread wavefront traversal in space-time blocks.
    Front\-lines are updated sequentially.]{
    \centering
        \includegraphics[clip=true, trim = 0.0cm 0.0cm 0.0cm 0.0cm ,width=3.5cm]{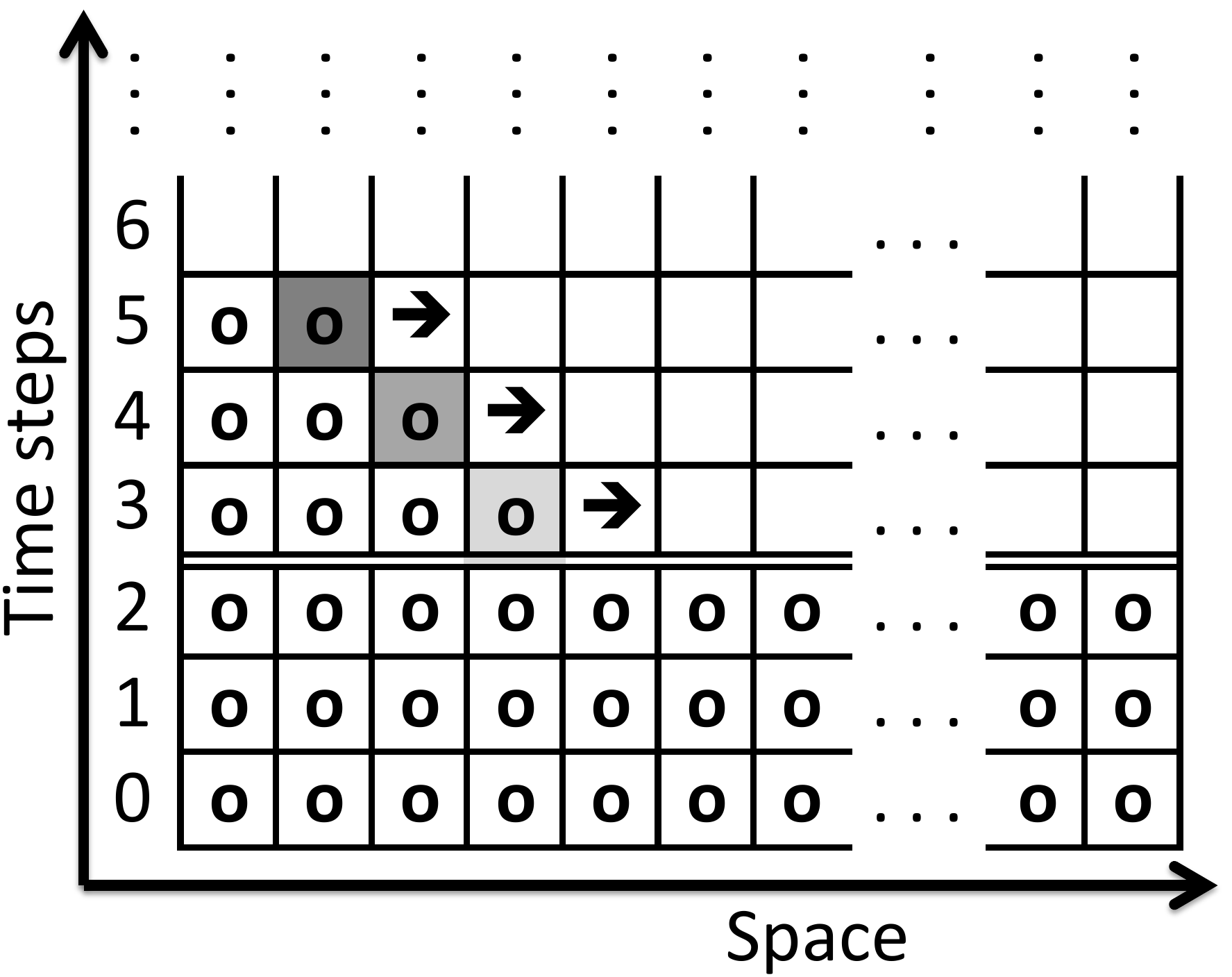}
    \label{fig:1_core_wavefront_3pt_1d}
    }
     \quad
    \subfloat[Multi-thread wavefront traversal in space-time blocks.
    One frontline update per thread, with thread synchronization after each time step update in the frontline.]{
    \centering
        \includegraphics[clip=true, trim = 0.0cm 0.0cm 0.00cm 0.0cm ,width=4.4cm]{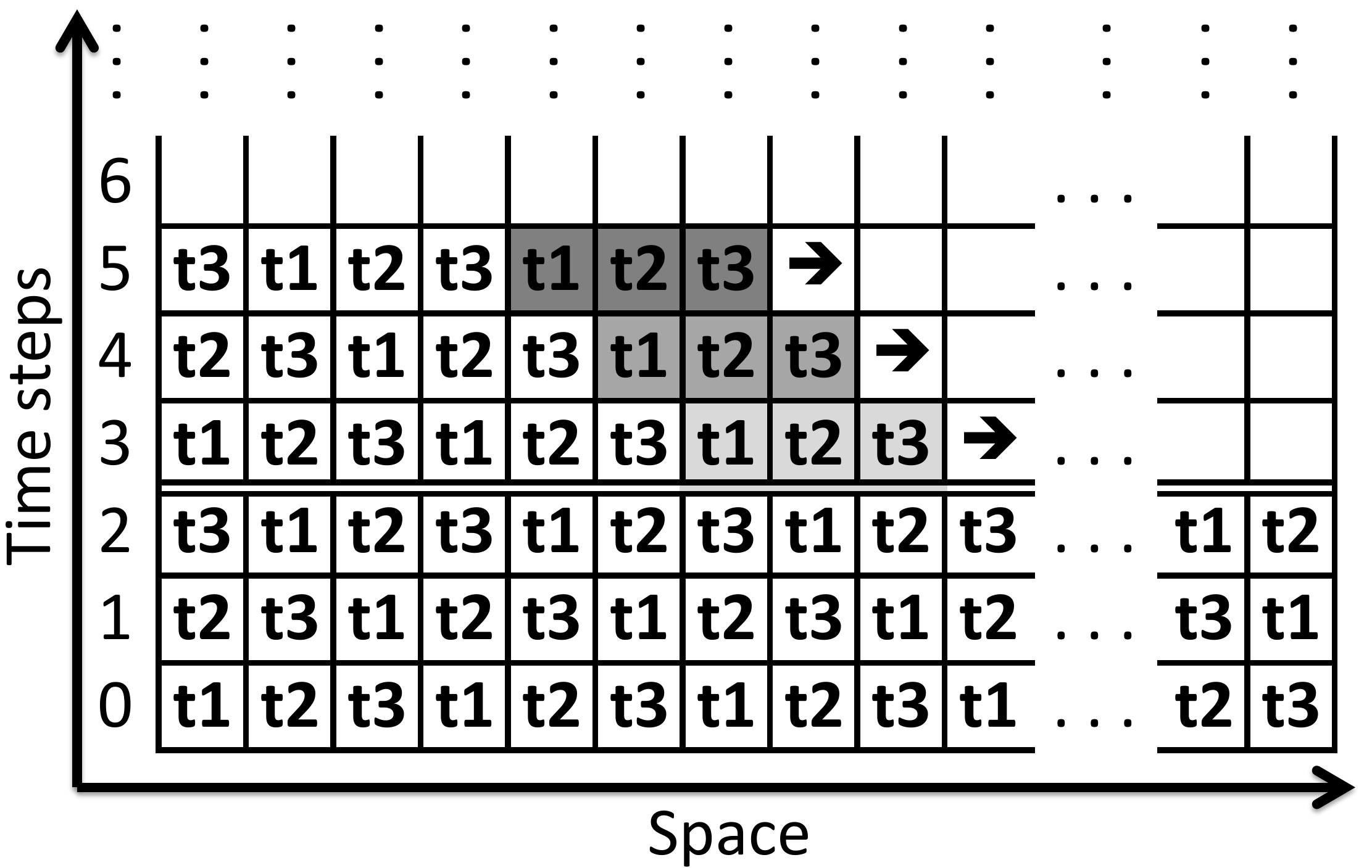}
    \label{fig:multi_core_wavefront_3pt_1d}
    }
    \caption{Different stencil update approaches for the 3-point stencil in one dimension.
    Fading gray boxes represent the last three updates.} 
\end{figure*}

Figure~\ref{fig:naive_3pt_1d} shows the na\"{i}ve update order of a 3-point 
stencil in one dimension (see the pseudo-code in
Fig.~\ref{fig:3_pt_stencil_1d}). The fading gray
color represents recently updated grid points, with the darkest
assigned to the most recent update.  The three ``upward pointing'' arrows  in time steps three and four display the
data dependency at each grid point, which is important to consider at
temporal blocking optimizations.  Wavefront temporal blocking is a
well known technique in the literature 
\cite{Lamport:1974,Wonnacott845979,Strzodka:2011:CAT}.  
Compared to the na\"{i}ve approach, the grid points
update order maximizes the reuse of the most recently visited grid
points while respecting the data dependencies.
Figure~\ref{fig:1_core_wavefront_3pt_1d} shows the basic idea of 
wavefront temporal blocking for the 3-point stencil in one dimension.
A wavefront line (``frontline'') traverses a space-time block in the
direction of the arrows.  The slope $S$ of the frontline is
determined by the radius of the stencil operator $R$ , where
higher order (i.e., longer range) stencils require a smaller frontline slope to respect the
data dependency ($S = -1/R$). For example, the 3-point stencil has
$S=-1$.  The data in the frontline has to fit in the cache memory
(along with the surrounding grid points touched by the stencil
operator) to achieve the desired cache memory data reuse in the
wavefront approach.  The time dimension is blocked in a size that
allows the frontline to fit in the desired cache.  For example, three
time steps are blocked in Figure~\ref{fig:1_core_wavefront_3pt_1d} to
illustrate the idea.


The wavefront can be executed sequentially (``1-thread
wavefront'').  On multi-core systems one can perform 1-thread
wavefronts on separate space-time tiles, with no need (nor use) for
any cache sharing among the threads. This approach was demonstrated in
detail in \cite{Strzodka:2011:CAT}.

An explicitly multi-core aware wavefront scheme leveraging the shared
cache among the threads was proposed in \cite{wellein5254211} (Similar concept in exploiting parallelism within the cache block was recently performed by Shrestha \textit{et al.}~\cite{shrestha2014jagged}).  In \cite{wellein5254211} approach the frontline update is
pipelined over a group of threads sharing an outer-level cache.  This
has the advantage of reducing the memory bandwidth pressure
and the total cache size requirement compared to the 1-thread
wavefront.  Figure~\ref{fig:multi_core_wavefront_3pt_1d} shows the
multi-thread wavefront variant used in this paper:
at each time step in the
multi-frontline update, each thread performs an update of one or
more grid points before proceeding to the next time step.  
All threads have to update the same number of
frontlines for load balancing, and they must be synchronized 
after each time step update to ensure 
correctness. A global barrier is the simplest solution for this, 
but a relaxed synchronization scheme may result in better performance
if the workload per thread is small~\cite{wittmann10}.

\begin{SCfigure*}
\centering
\caption{Diamond tiling on a one-dimensional space grid, with arrows
  representing inter-tile data dependencies.  The number of diamond
  tiles per row represents the maximum attainable concurrency,
  as the tiles in the row can be executed independently of each other.}
\includegraphics[width=7cm]{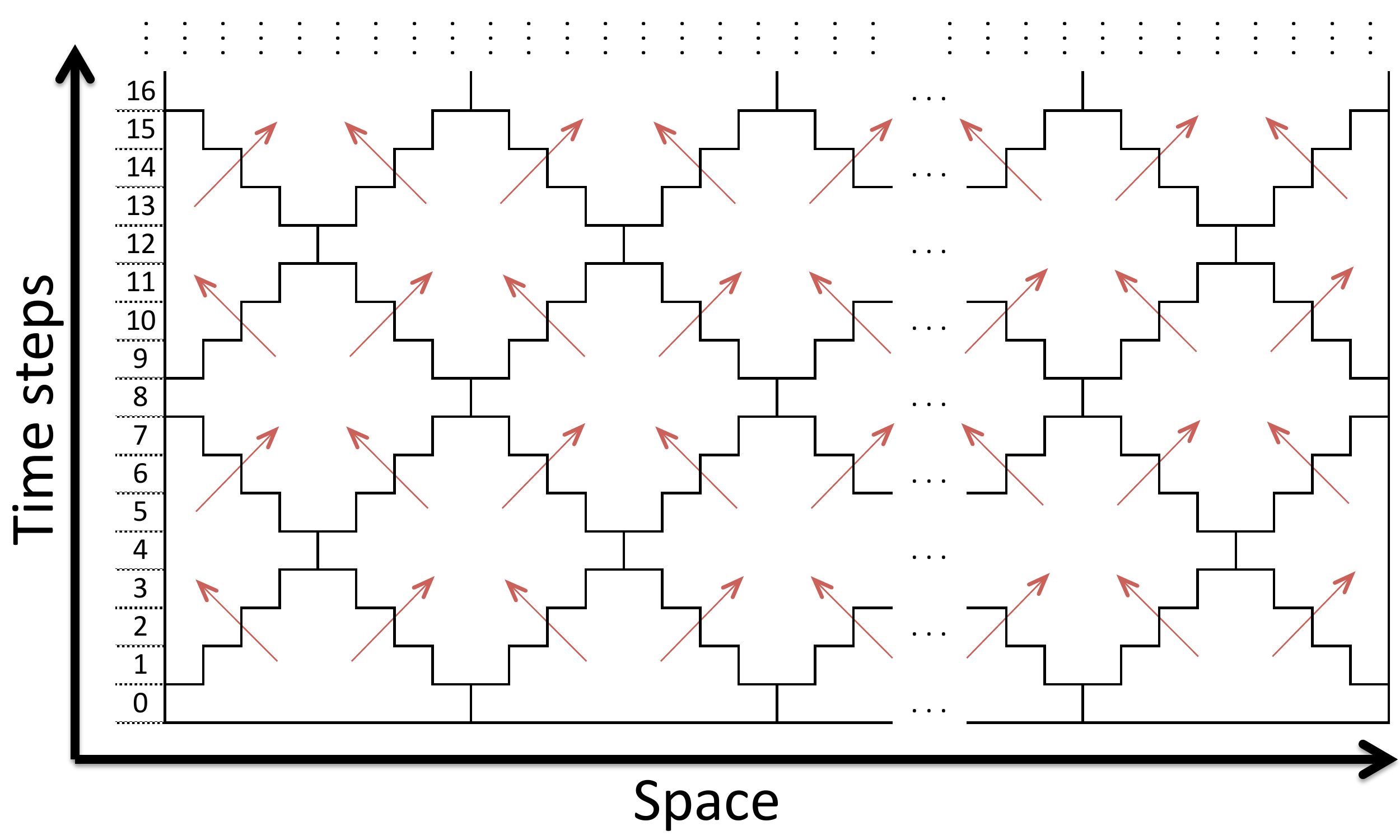}
\label{fig:diamond_tiling_1d}
\end{SCfigure*}

\subsection{Diamond tiling}
Diamond tiling has received much attention in recent years.
Figure~\ref{fig:diamond_tiling_1d} shows the basic idea for the
one-dimensional 3-point stencil. Arrows represent the data dependency
across the diamond tiles.  Diamond tiles that start at the same time
step compose a ``row of diamonds'', in which the diamond tiles are
independent of each other.  Each interior diamond tile has a data
dependency on the two diamond tiles that share edges with it in the
lower row of diamonds (``parents'').  Half-diamond tiles at the boundaries of
the spatial domain have only one parent.  The slope of the tile edges
depends on the stencil radius, where $S=\pm 1/R$.  We consider atomic
update of the diamond tiles in this work, i.e., a tile does not
perform synchronization with other tiles until it is completely updated.  Several
advantages of diamond tiling make it favorable in shared memory
systems: it maximizes the data reuse of the loaded data block~\cite{orozco2009ICPP},
has low synchronization requirements, allows concurrent start-up 
in updating the diamond tiles, and uses a unified tile shape, 
which simplifies the implementation.

\subsection{1-thread wavefront diamond blocking}

A multi-dimensional tiling algorithm called ``cache accurate time skewing'' (CATS) was proposed at~\cite{Strzodka:2011:CAT}. They use
space-time tiles larger than the cache size and employ wavefront
temporal blocking inside the tiles to improve the data reuse in the
cache memory (similar to~\cite{nguyen18993, wellein5254211} in combining wavefront with large tiles). In CATS, the leading space dimension is not tiled and 
left for vectorization, while one other space dimension is used for wavefront and blocking (When running at more than one dimensional grid). 

We are interested in the CATS2 variant, where ``2'' stands for diamond
tiling in one dimension plus wavefront in another dimension. It is
suitable for three-dimensional problems and for distributed memory 
parallelization, which will be described
in Section~\ref{approach}.  Figure~\ref{fig:1wd_2d} shows one extruded
diamond block of the CATS2 algorithm applied to a three-dimensional
problem.  The wavefront traverses along the $z$ dimension. The diamond 
tiling is performed across the $y$ dimension, allowing the diamond tile to perform spatial blocking to satisfy the layer condition. The
$x$ dimension is left intact  for
efficient hardware data prefetching and minimum TLB misses.
Hence, each grid point in Fig.~\ref{fig:1wd_2d} extends along
the full $x$ range.

\section{Approach: multi-thread wavefront diamond blocking}
\label{approach}

The implementation of our approach can be found in~\cite{malas_girih_2014}. We use it to produce the results of this work.

\subsection{Tiling and threading scheme}

The CATS2 algorithm has the advantage of efficiently utilizing the
performance of multi-core processors with minimal thread
synchronization and very efficient data reuse.  However, we believe
that two particular aspects of CATS can be potentially improved.

First, CATS2 relies on a large domain size in the diamond tiling
dimension to have sufficient concurrency for the available threads.
If this condition cannot be met, \cite{Strzodka:2011:CAT} proposes reverting to
CATS1, which uses wavefront traversal in the same dimension of
space-time parallelogram tiles.  This prevents the CATS algorithm from taking advantage of
diamond tiling as an efficient domain decomposition strategy in
distributed memory.
They also propose using higher-dimensional tiling to extract parallelism for many-core processors, which can increase the code complexity.
Moreover, with the emergence of many-core architectures like the Intel
Xeon Phi, it would be difficult to find sufficient concurrency for
60--240 threads in a reasonable grid size.

Second, no cache sharing among threads is assumed in the CATS2
algorithm, so each thread requires space in the cache memory and
bandwidth from main memory for its own use.  As a result,
memory-starved stencil computations run out of cache and memory
bandwidth, as will be shown in the results section.  Moreover, it is
unclear whether the cache size and memory bandwidth per thread as seen
in contemporary multi-core designs will be available in future
architectures. For example, the Intel Xeon Phi has 128\KiB\
and 8\KiB\ cache per thread in the L2 and L1 caches, respectively,
and it achieves only about 3\GBS\ per core of memory bandwidth in full 
saturation. A temporal blocking scheme should be flexible enough
to accommodate such hardware limitations if required.

\begin{figure*}[tbp]
    \centering
    \subfloat[Diamond tiling with a single-thread wavefront 
    (similar to CATS2) in a three-dimensional space grid using 
    a single frontline.]{
        \centering
        \includegraphics[width=4.7cm]{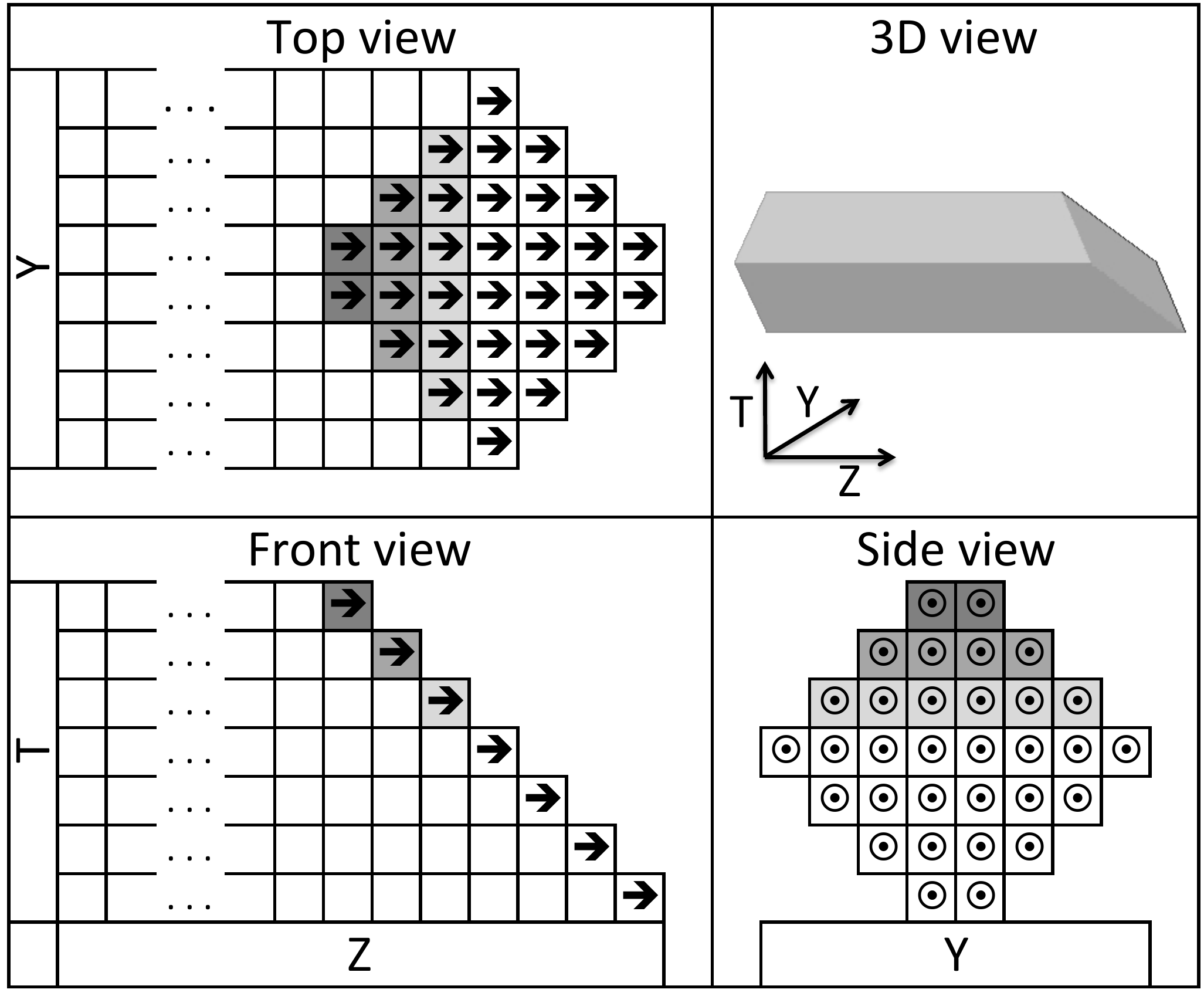}
        \label{fig:1wd_2d}
    }
    \quad
    \subfloat[Diamond tiling with a multi-thread 
	wavefront (shown here with two threads) in a three-dimensional 
	space grid, using two frontlines per thread.]{
        \centering
        \includegraphics[width=6.6cm]{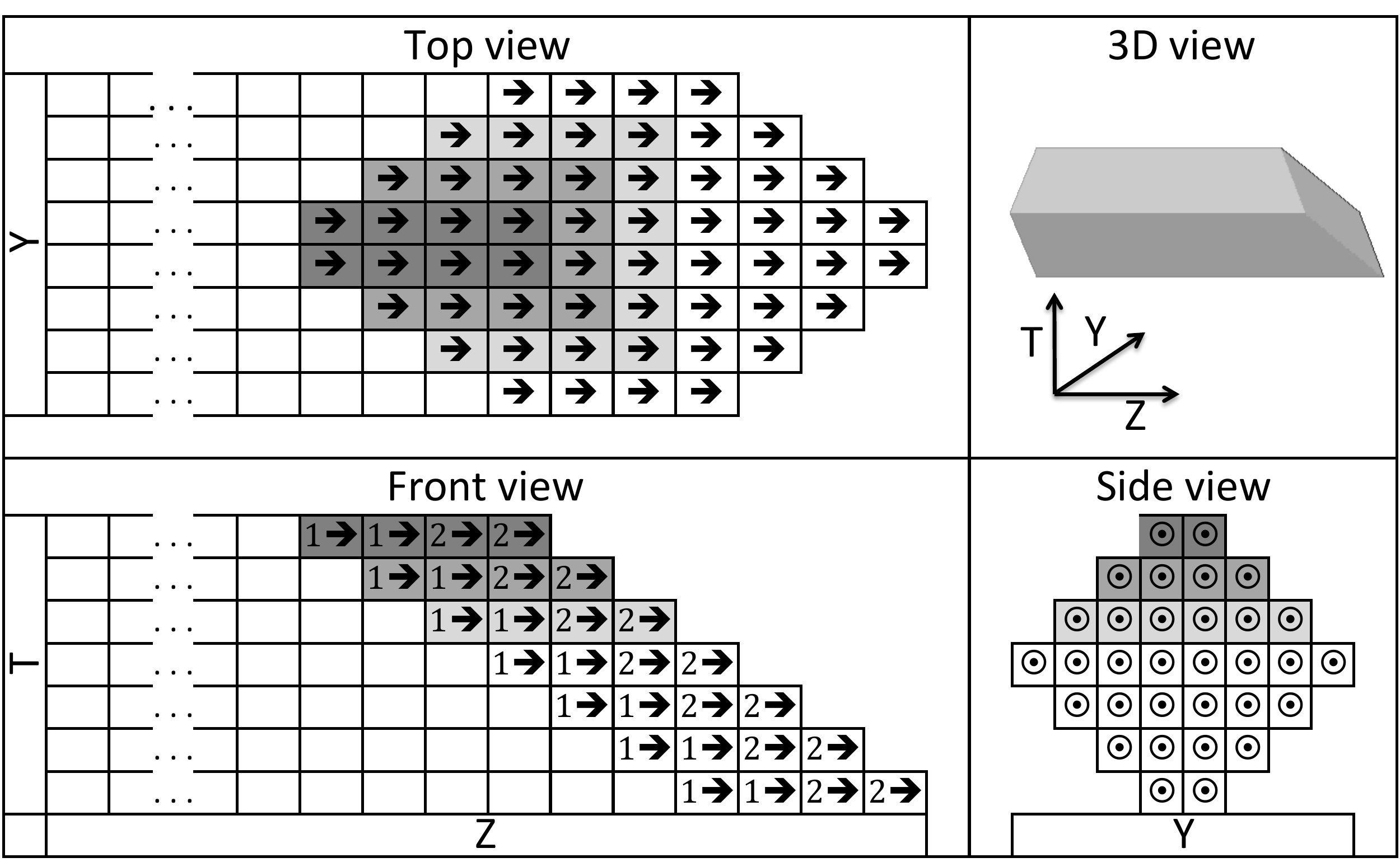}
        \label{fig:mwd_2d}
    }
    \caption{Diamond tiling with wavefront temporal blocking.}
\end{figure*}

We propose an approach that reduces the cache size and
memory bandwidth requirements and reduces the concurrency limitations
while retaining the advantages of CATS2.  The basic idea is to use the
multi-threaded wavefront temporal blocking proposed in~\cite{wellein5254211} in place of the CATS2 1-thread wavefront.  
We call our concept multi-threaded wavefront diamond
blocking (MWD), where CATS2 is a special case with 1-thread wavefront
diamond blocking (1WD). Figure~\ref{fig:mwd_2d} shows one block of the
MWD algorithm.  
Load-balancing is important here as the number of stencil updates in the
 frontline varies across time steps according to the diamond width.
The multi-thread wavefront updates are performed in the same manner as 
described in Sect.~\ref{sec:wavefront}.  The parallelization
of each time step of the multi-frontline update ensures load balancing
among the threads.

Threads are assigned to the extruded diamond in groups (``thread
groups'').
For example, two-threads wavefront diamond blocking (2WD) is used in
Figure~\ref{fig:mwd_2d}.  Multiple thread groups can run concurrently,
working on different diamonds tiles and respecting inter-diamond
dependencies.

One advantage of MWD over CATS2 is controllable cache sharing among
the threads.  When multiple threads update an extruded diamond
together, they share the memory bandwidth and the cache memory block,
reducing the pressure on these resources.  Moreover, less concurrency
is required across multiple diamonds (i.e., in the $y$ dimension),
since some concurrency is moved to the wavefront dimension.  This allows
for smaller domain sizes without sacrificing concurrency.

Threads can be scheduled to the extruded diamonds in a variety of
ways.  Orozco \textit{et al.}~\cite{orozco2009ICPP} use global synchronization
after each row of diamonds update to respect the inter-diamond
dependencies. The diamond tiles in~\cite{Strzodka:2011:CAT} are preassigned 
to threads before starting the stencil computations,
taking the inter-diamond data dependency during tile updates into
account. These approaches are sufficient to avoid idling threads as
long as the workload is balanced.  Workload variation can result from
domain boundary handling: In this work, diamond tiles at the boundary
of the subdomain exchange data and synchronize with neighbor
processes.  This causes load imbalance in processing diamond tiles,
which varies according to the used network interconnect.  To resolve
this issue, we schedule the diamond tiles dynamically to the thread
groups. A FIFO queue maintains a list of available tiles for update.
 When a thread has completed updating a tile, it pushes its
dependent diamond tile(s) to the queue if those tiles have no other
unmet dependencies.  ``Pop'' operations are performed to assign available tiles to thread groups.
The FIFO queue is protected from concurrent updates
by an OpenMP critical region.  Since the queue updates are
performed infrequently, the synchronization overhead is negligible.

Proper choices of the diamond tile size and the number of
multi-frontline updates are crucial for good performance.
In particular, maximizing the diamond tile size in the L3 cache increases the data reuse in the L3 cache. The diamond tile size in \cite{Strzodka:2011:CAT} is computed  based on the processor's cache memory size and the stencil
operator specification.  This approach does not guarantee best
cache utilization to maximize the data reuse.  The optimal
cache block size can vary based on the cache memory architectural
features, such as associativity, and the data access patterns of the
used stencil operator.  Auto-tuning is used in this work to select the
diamond tile size and number of multi-frontline wavefront updates.
The parameter search space is narrowed down to diamond tiles that fit
within a predefined cache size range.  The cache block size is
computed based on diamond size, number of wavefronts, grid size, and
stencil type, as will be described in Section~\ref{sec:mwd_analysis}.
Several constraints are considered in selecting valid cache block size, 
for example, having sufficient concurrency and integer number of diamond tiles in each row of diamond tiles.


\subsection{Distributed-memory parallelization}\label{sec:approach:distmem}

Parallelizing stencil computations over distributed memory nodes is
quite straightforward if no temporal blocking is involved. Each time
step update is followed by halo data communication.
In such a bulk-synchronous scheme, strong scalability is naturally
limited by data transfer overhead.
A partial remedy is provided by the halo-first update scheme, in which 
domain boundaries are updated first, and then asynchronous 
message passing is performed while updating the bulk of the
domain.

\begin{figure*}[tbp]
\centering
\includegraphics[width=9cm]{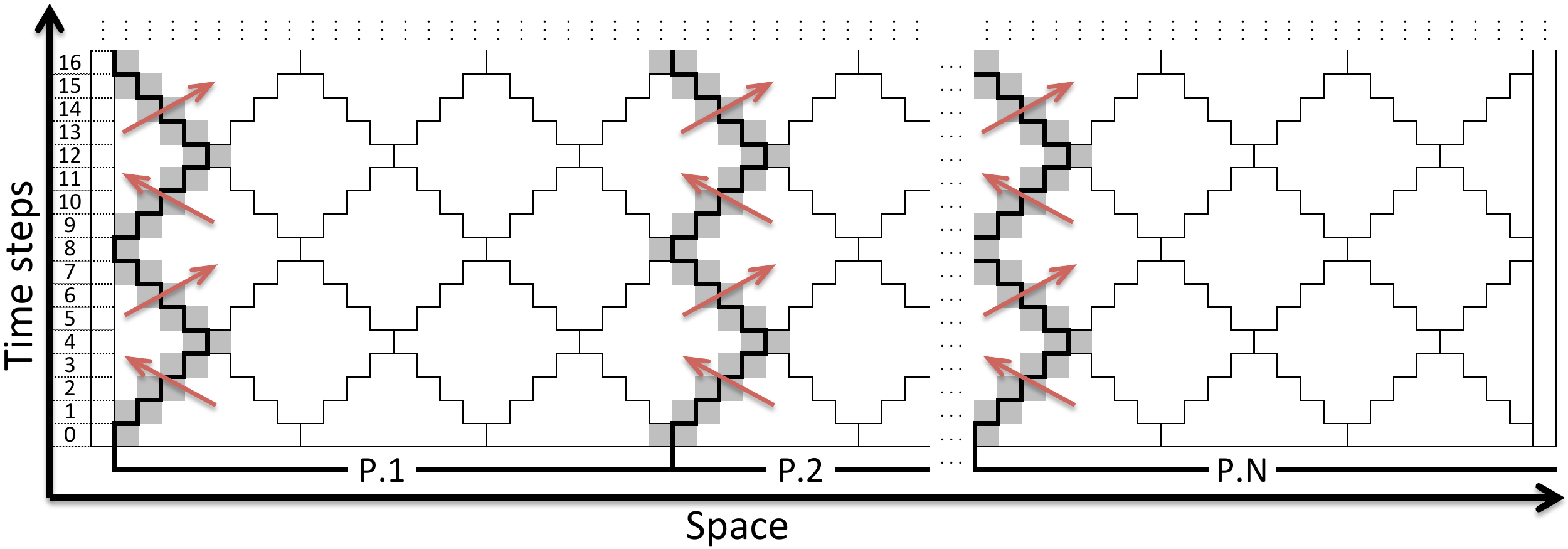}
  \caption{Distributed memory parallelization with diamond tiling for a
    one-dimensional space grid. Arrows represent the data dependencies
    across subdomains. The leftmost column of half diamonds is
    assigned to the rightmost process to achieve load balance in
    computation and communication.}
\label{fig:diamond_tiling_distributed_1d}
\end{figure*}

Distributed memory parallelization can be combined with diamond tiling
as shown in Figure~\ref{fig:diamond_tiling_distributed_1d}.  The
arrows represent the data dependencies across subdomains, and the same
number of adjacent tiles is assigned to each process except
the rightmost one (largest $y$ coordinate).  To
maintain load balance in terms of computation and communication, the
leftmost half diamond tile is assigned to the rightmost process.
Regular diamond tiles are used at the boundary of subdomains, with the
difference of performing communication before and after the
tile update.  Thread groups handling boundary diamond tiles are blocked until their MPI communication is complete.  Extra delay can occur if no 
thread group is updating the
diamond tile at the other end of the communication.  Adding priority
in scheduling the tiles at the boundary to thread groups can alleviate
this issue, which is left for future work.

Since domain decomposition is performed along the middle space
dimension ($y$), the boundary data to be communicated does not reside
in contiguous memory locations. User-defined strided MPI data
types are not efficient in our multi-threaded implementation, as 
MPI implementations handle the required packing/unpacking operations
purely sequentially. We use explicit multi-threaded halo data
packing/unpacking to resolve this issue.

Diamond tiling offers several advantages in distributed memory
parallelization. The tessellation of the diamond tiles allow using a unified tile structure everywhere. It also allows maximum stencil updates 
in space-time without relying on exchanging boundaries with neighbor 
processes after each grid sweep. Finally, there is a natural overlap of 
computation with communication. Communication does not block all threads, 
and no thread has to be sacrificed for asynchronous communication.  
Threads can handle communication or perform stencil updates as needed.

\subsection{Wavefront diamond blocking cache block size requirement}\label{sec:mwd_analysis}

Four parameters are used to calculate the cache block size of the
wavefront: the diamond width $D_w$ in the $y$ axis, the number of
additional wavefront frontlines $N_{F}$, the number of bytes in the
$x$ axis $N_{xb}$, and the stencil operator properties (see
below). The wavefront width $W_w$ is determined by the diamond width
and the number of frontlines: $W_w=D_w+N_{F}-1$.  The stencil operator
properties include the stencil radius. for
example, the 7- and 25-point stencils have $R\! =\! 1$ and $R\! =\!
4$, respectively.  They also include the number of additional domain-sized streams used in the stencil operator, $N_D$.  For example, the 7-point stencil
with constant coefficients uses two domain-sized streams (Jacobi style update).  
The 7-point stencil with variable coefficients uses seven
additional domain-sized streams to hold the coefficients.
Total number of bytes required in the cache block for a stencil 
with $R\! =\! 1$, with some approximations, is:
\bq\label{eq:7pt_cache_blk_size}
N_{xb} \cdot \left[N_D\cdot \left(\frac{D_w^2}{2} + D_w\cdot N_{F}\right) 
  + 2 \cdot(D_w + W_w)\right] \eos
\eq
Here, $N_{xb}$ is the size
of the ``untiled'' leading dimension, and ${D_w^2}/{2} + D_w\cdot N_{F}$ is the
diamond area in the $y$-$z$ plane as shown in the top
view of Figures~\ref{fig:1wd_2d} and \ref{fig:mwd_2d}.
The perimeter of the rectangle containing the cache block in the $y$-$z$ plane
is $2 \cdot(D_w + W_w)$.
For example, we have $D_w=8$ and $N_{F}=0$ in Figure~\ref{fig:1wd_2d}, so 
$W_w = 8-1+0 = 7$ and the total block size is $N_{xb} \cdot ((8^2/2 +
8\cdot 0) + 2 \cdot(8 + 7)) = 70\cdot N_{xb}$~\bytes. 
Figure~\ref{fig:mwd_2d} differs with $N_{F}=3$, so $W_w = 8-1+3
= 10$ and the cache block size is $N_{xb} \cdot ((\frac{8^2}{2} + 8\cdot
3) + 2 \cdot(8 + 10)) = 92\cdot N_{xb}$~\bytes.

Higher-order stencils (with $R>1$) have a steeper frontline slope 
in the wavefront to satisfy the data dependency across time steps.
This results in different wavefront lengths ($W_w= D_w - 2\cdot
R + N_{F}+1$).  Also the required cache block size changes as follows:
\bq\label{eq:cache_blk_size}
N_{xb} \cdot \left[N_D\cdot D_w\cdot \left(\frac{D_w}{2} - R + N_{F} + 1\right) 
  + 2 \cdot R \cdot (D_w + W_w)\right]\eos
\eq

\begin{SCfigure*}
    \centering
    \caption{Model of MWD cache block size requirements per thread 
      group at different diamond tile sizes.
      Each data set corresponds to a different grid size.}
    \subfloat[7-point stencil with  constant coefficients stencil.]{
        \centering
        \includegraphics[width=4.5cm]{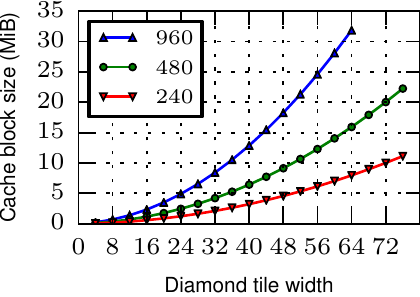}
    }
    \enskip
    \subfloat[25-point stencil with variable coefficients stencil.]{
        \centering
        \includegraphics[width=4.3cm]{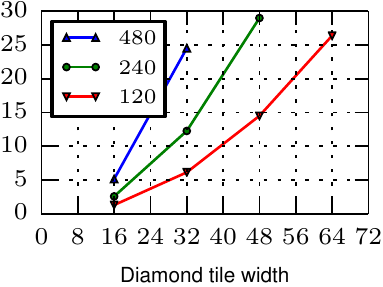}
    }
    \label{fig:all_cache_req}
\end{SCfigure*}

Examples of wavefront cache block size requirements for the 7-point
constant-coefficient and the 25-point variable-coefficient stencils
are shown in Figure~\ref{fig:all_cache_req}.  Considering that each
thread group requires a separate cache block, the 25-point stencil
with variable coefficients needs a lot of cache on modern multi-core
processors, even at the smallest diamond tile size.  When using the
1WD algorithm on a ten-core CPU, 50\,\MiB\ of cache memory are
required, which is beyond the capacity of current designs.  On the
other hand, the 7-point stencil with constant coefficients can be used
with a sufficiently large diamond tile size and still fulfill the
cache demand even with 1WD. Since each data set in
Fig.~\ref{fig:all_cache_req} represents the cache size for only one
thread group, it is evident that, all other parameters being equal,
MWD allows for larger tile sizes than 1WD.  

\section{Results}\label{sec:results}

In this section we compare the performance of spatially blocked code
with our MWD variants.  On the ten-core Intel Ivy
Bridge processor, four thread group sizes are investigated: 1WD, 2WD,
5WD, and 10WD.  Since we perform parameter auto-tuning and dynamic
scheduling of diamond tiles to threads, our 1WD approach is an
improved version of CATS2.

We show performance results for those stencils that were analyzed in
the previous sections and shown to be strongly memory bound so that
temporal blocking is a viable optimization: 7-point constant-coefficient (Fig.~\ref{fig:7_pt_stencil_3d}), 7-point variable-coefficient (Fig.\ref{fig:7_pt_var_stencil_3d}), and 25-point variable-coefficient (Fig.~\ref{fig:25_pt_var_stencil_3d}).

Three sets of results are presented:
OpenMP thread scaling performance on a single socket
(Sect.~\ref{sec:intrasocket}), increasing cubic grid size performance
in a single socket (Sect.~\ref{sec:perf_vs_size}), and distributed
memory strong scaling performance (Sect.~\ref{sec:distmem}).  For all
performance results we chose
the grid size that fits in 32\,\GiB\ of memory, which is the
memory size per socket in the benchmark system.

\subsection{Thread scaling performance}\label{sec:intrasocket}

Although we present scaling runs from 1 to 10 threads on an Intel
Ivy Bridge socket, with fixed grid size, auto-tuning is
performed at 10 threads only.  To further improve the results, a brute-force parameters search is perform at the 10 threads experiments for better 
results.  For each of the thread
scaling results, the same diamond width and number of frontlines of
the corresponding 10-threads experiment is used.

\subsubsection{7-point stencil with constant coefficients}\label{sec:7pt_const}

\begin{figure*}[tbp]
    \newcommand*{\sfwidth}{3.8cm}
    \centering
    \subfloat[7-point constant-coefficient stencil performance.]{
        \centering
        \includegraphics[width=4.2cm]{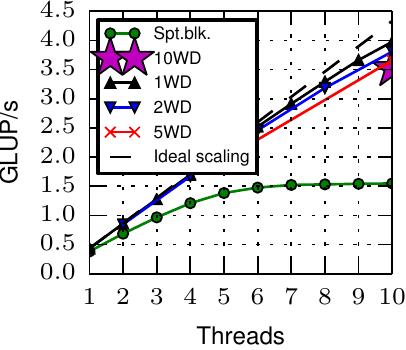}
        \label{fig:7_pt_const_all_methods_perf}
    }
    \enskip
    \subfloat[7-point variable-coefficient stencil performance.]{
        \centering
        \includegraphics[width=\sfwidth]{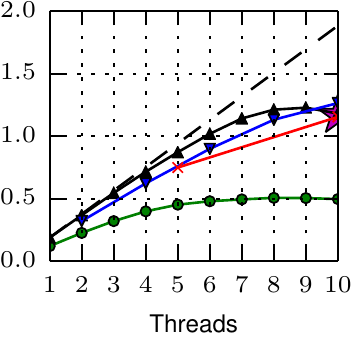}
        \label{fig:7_pt_var_all_methods_perf}
    }
    \enskip
    \subfloat[25-point variable-coefficient stencil performance.]{
        \centering
        \includegraphics[width=\sfwidth]{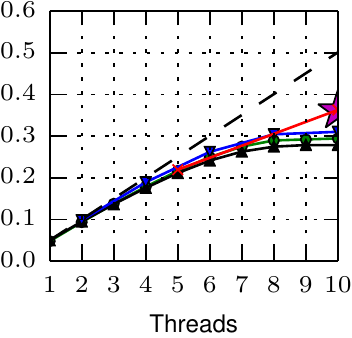}
        \label{fig:25_pt_var_all_methods_perf}
    }    

    \subfloat[7-point constant-coefficient stencil measured memory bandwidth.]{
        \centering
        \includegraphics[width=4.2cm]{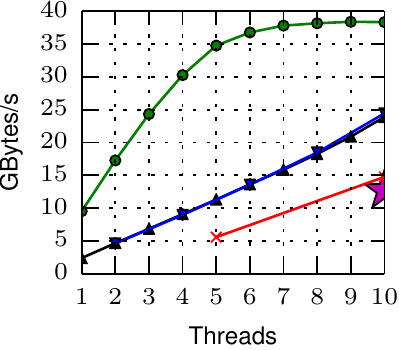}
        \label{fig:7_pt_const_all_methods_bw}
    }
    \enskip
    \subfloat[7-point variable-coefficient stencil measured memory bandwidth.]{
        \centering
        \includegraphics[width=\sfwidth]{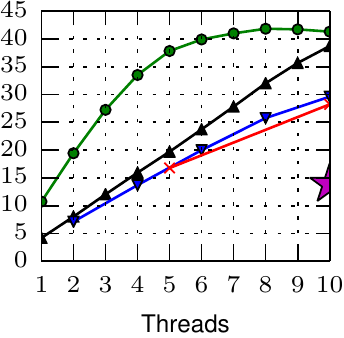}
        \label{fig:7_pt_var_all_methods_bw}
    }
    \enskip
    \subfloat[25-point variable-coefficient stencil measured memory bandwidth.]{
        \centering
        \includegraphics[width=\sfwidth]{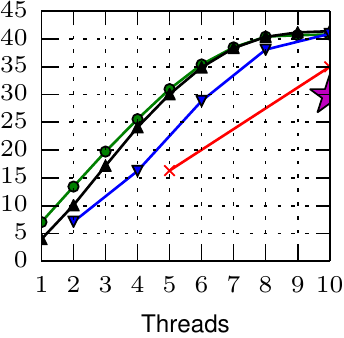}
        \label{fig:25_pt_var_all_methods_bw}
    }
    \caption{Performance (top row) and memory bandwidth (bottom row) 
      scaling vs.\ number of threads of three stencil types (columns) on a 10-core Intel 
      Ivy Bridge socket. Problem sizes: $960^3$, $680^3$, and $480^3$
      for 7-point constant-coefficient, 7-point variable-coefficient,and 25-point variable-coefficient, respectively.
      Ideal scaling is based on 1 thread 1WD performance at the three stencil types.}
\end{figure*}

All thread scaling results are shown in Figures~\ref{fig:7_pt_const_all_methods_perf} and ~\ref{fig:7_pt_const_all_methods_bw}.

As shown in Figure~\ref{fig:7_pt_const_all_methods_perf}, 1WD
  achieves a 2.6$\times$ speedup compared to spatially blocked code, with
  about 92\% threads parallel efficiency and 95\% of the in-L3 
  performance (see Fig.~\ref{fig:7_pt_const_inL3}) using 10 cores.
2WD, 5WD, and 10WD achieve 4\%, 8\%, and 12\% less performance
  compared to 1WD, respectively.
  Since all MWD variants are decoupled from main memory, the core
  performance and the intra-cache data transfers become the bottleneck.  
  The 1WD implementation has some advantage here since 
  it does not require OpenMP parallelism in the wavefront
  update, and the OpenMP barrier overhead
  increases at larger thread groups.  

In terms of memory bandwidth usage, as shown in
  Figure~\ref{fig:7_pt_const_all_methods_bw}, 1WD, 2WD, 5WD, and 10WD
  save about 40\%, 39\%, 63\%, and 68\% of the memory
  bandwidth, respectively.
10WD requires roughly 47\% less memory bandwidth compared to
  1WD, which makes it a better candidate on future, more
  bandwidth-starved architectures: Since 10WD uses 32\% of the
  available memory bandwidth, it can theoretically maintain the same
  performance on processors with a bandwidth that is a factor of
  three lower than on the benchmark system (assuming the same core
  performance). This result also raises expectations of 10WD showing
  more benefit or stencils with a higher code balance (see below).

In Table~\ref{table:7_pt_const_all_methods_energy} we show power
dissipation and energy to solution measurements taken with \lperfctr\
(which uses the RAPL facility available in modern Intel
processors). When considering aggregate DRAM and processor power, 10WD
draws 13\% less power compared to 1WD, which is mostly due to reduced
DRAM bandwidth usage.  Due to its slightly lower performance, 10WD can
still not achieve significant energy savings compared to 1WD. However,
future architectures are expected to show a much larger contribution
to overall power from memory transfers, which will make MWD approaches
more attractive also in terms of energy consumption. Note also that
all variants with temporal blocking dissipate more power on the CPU
but less in the DRAM compared to spatial blocking. Together with the
much improved time to solution this results in considerable
energy savings, even when running the spatially blocked version 
with only six cores, i.e., at the performance saturation point.
For code with saturating performance characteristics, this is
the point with minimal energy to solution~\cite{CPE:CPE3180}.

A more detailed energy consumption analysis is left to future work.

\begin{table}[tbp]
\centering
\begin{tabular}{|r|r||c|c|c|c|c|c|}
\hline
 & Method & \multicolumn{ 2}{|c|}{Spt. Blk.} & 1WD & 2WD & 5WD & 10WD \\
\cline{2-8}
 & Threads & 10 & 6 & 10 & 10 & 10 & 10 \\
\hline
 Power & CPU & 52.9 & 44.1 & 58.5 & 59.0 & 57.4 & 56.6 \\
\cline{2-8}
 [W] & DRAM & 42.5 & 39.8 & 32.6 & 33.5 & 24.4 & 22.4 \\
\cline{2-8}
 & Total & 95.4 & 83.9 & 91.1 & 92.4 & 81.7 & 79.0 \\
\hline
 Energy & CPU & 33.9 & 29.6 & 14.7 & 15.5 & 15.8 & 16.2 \\
\cline{2-8}
[J] & DRAM & 27.3 & 26.8 &  8.2 &  8.8 &  6.7 &  6.4 \\
\cline{2-8}
 & Total & 61.2 & 56.4 & 22.9 & 24.4 & 22.5 & 22.7 \\
\hline
\end{tabular}
\caption{Power dissipation and energy to solution for the 7-point stencil 
with constant coefficients on a 10-core Intel Ivy Bridge socket at a grid 
size of $960^3$.}
\label{table:7_pt_const_all_methods_energy}
\end{table}

\subsubsection{7-point stencil with variable coefficients}
All thread scaling results are shown in Figures~\ref{fig:7_pt_var_all_methods_perf} and \ref{fig:7_pt_var_all_methods_bw}.

2WD achieves the best performance, as shown in
 Figure~\ref{fig:7_pt_var_all_methods_perf}, with a speedup of about
 2.55$\times$ compared to spatially blocked code.
The performance saturation and decay of 1WD beyond 7 threads deserves
an explanation.  Since the memory bandwidth scales almost linearly up
to 10 threads, one must conclude that the code balance
increases. Experiments with half the inner domain size (i.e.,
340$\times$680$\times$680) show that scaling continues up to 10 threads,
topping at about 1.5\,\GLUPS. It turns out that for 1WD the required
cache block size at ten threads is very close to the available cache size at the selected diamond size (see Sec.~\ref{sec:thread_scaling_diamond_size_analysis} for details). 
This problem can be remedied by split tiling in the leading dimension,
which we leave for future work.

2WD, 5MWD, and 10WD decouple from main memory, saving up to about two thirds
of the memory bandwidth in case of 10WD. 

Similar memory bandwidth usage of 2WD and 5WD is measured in
Figure~\ref{fig:7_pt_var_all_methods_bw} because the same diamond tile
size ($D_w=8$) is selected by the auto-tuner, compared to 10WD that
has a larger diamond tile size ($D_w=20$).  5WD has the same diamond
tile size as 2WD in this case because the next larger valid diamond
size would require more cache than is available.

\subsubsection{25-point stencil with variable coefficients}
All thread scaling results are presented in Figures~\ref{fig:25_pt_var_all_methods_perf} and \ref{fig:25_pt_var_all_methods_bw}.

Since this stencil has a very large code balance of $120\,\bytes/\LUP$, 
we expect larger thread groups to perform better.
The memory bandwidth of the spatial blocking, 1WD, and 2WD codes
cannot decouple from memory and saturate at eight threads, as shown in
Figure~\ref{fig:25_pt_var_all_methods_bw}.  On the other hand, 5WD and
10WD save roughly 12.5\% and 25.5\% of the memory bandwidth,
respectively, which allows them to have scalable performance as shown
in Figure~\ref{fig:25_pt_var_all_methods_perf}.
5WD and 10WD achieve about 1.23$\times$ speedup compared to the
spatially blocked code.

The 1WD algorithm achieves worse performance compared to spatially
blocked code due to cache capacity misses, since it requires 93\,\MiB\
of cache even at the smallest possible diamond tile size.


\subsubsection{Thread scaling diamond tile size analysis} \label{sec:thread_scaling_diamond_size_analysis}

The 7-point stencil with constant coefficients achieves the best performance with $D_w=12$ 
at 1WD and 2WD. 
Although 2WD has less cache size and memory bandwidth requirements than 1WD, it uses the same diamond width because 
the next larger diamond width ($D_w=16$) does not fit in the L3 cache. 
2WD requires 25.4\MiB\ cache block size at $D_w=16$ compared to 17.5\MiB\ at $D_w=12$, so $D_w=16$ can cause a lot of capacity misses to the main memory.
The best performance at 5WD and 10WD is achieved at $D_w=24$.
Figure~\ref{fig:7_pt_const_all_methods_bw} shows that the same memory bandwidth usage 
is obtained when the same diamond width is used.

The 7-point stencil with variable coefficients uses $D_w=8$ 
at 1WD, 2WD, and 5WD, and $D_w=20$ 
at 10WD.
1WD uses more memory bandwidth than 2WD and 5WD, although they have 
the same code balance. The best performance in 1WD is obtained at a diamond width that causes many capacity misses to main 
memory, because the selected cache block size is large. 
Although the next smaller diamond width ($D_w=4$) can avoid these 
cache misses in 1WD, it achieves lower performance 
due to  the higher code balance.
1WD has similar memory bandwidth usage to 2WD and 5WD at eight threads, 
where the cache block size does not cause many cache capacity misses.

The 25-point stencil with variable coefficients uses $D_w=16$ at 1WD, 2WD, 
and 5WD, and $D_w=32$ at 10WD. The cache block size at 1WD and 2WD is larger 
than the available cache size even with smallest possible diamond width 
($D_w=16$), causing the memory bandwidth to saturate in these cases.

\subsection{Performance at increasing grid size}\label{sec:perf_vs_size}

Here we compare the performance of problems with increasing cubic grid
size.  The grid size in each dimension is set to a multiple of 64.

Performance and memory bandwidth measurements are shown in Figures
\ref{fig:increasing_grid_size}.
In all cases with temporal blocking, but especially for the 7-point stencil with
variable coefficients, a gradual decline of performance with growing
grid size can be observed. This can be attributed to the same
cause as the bad thread scaling for 1WD beyond 7 threads in 
Fig.~\ref{fig:7_pt_const_all_methods_perf}.

\subsubsection{7-point stencil with constant coefficients}
\begin{figure*}[tbp]
    \centering
    \subfloat[7-point constant-coefficient stencil performance.]{
        \centering
        \includegraphics[width=4.1cm]{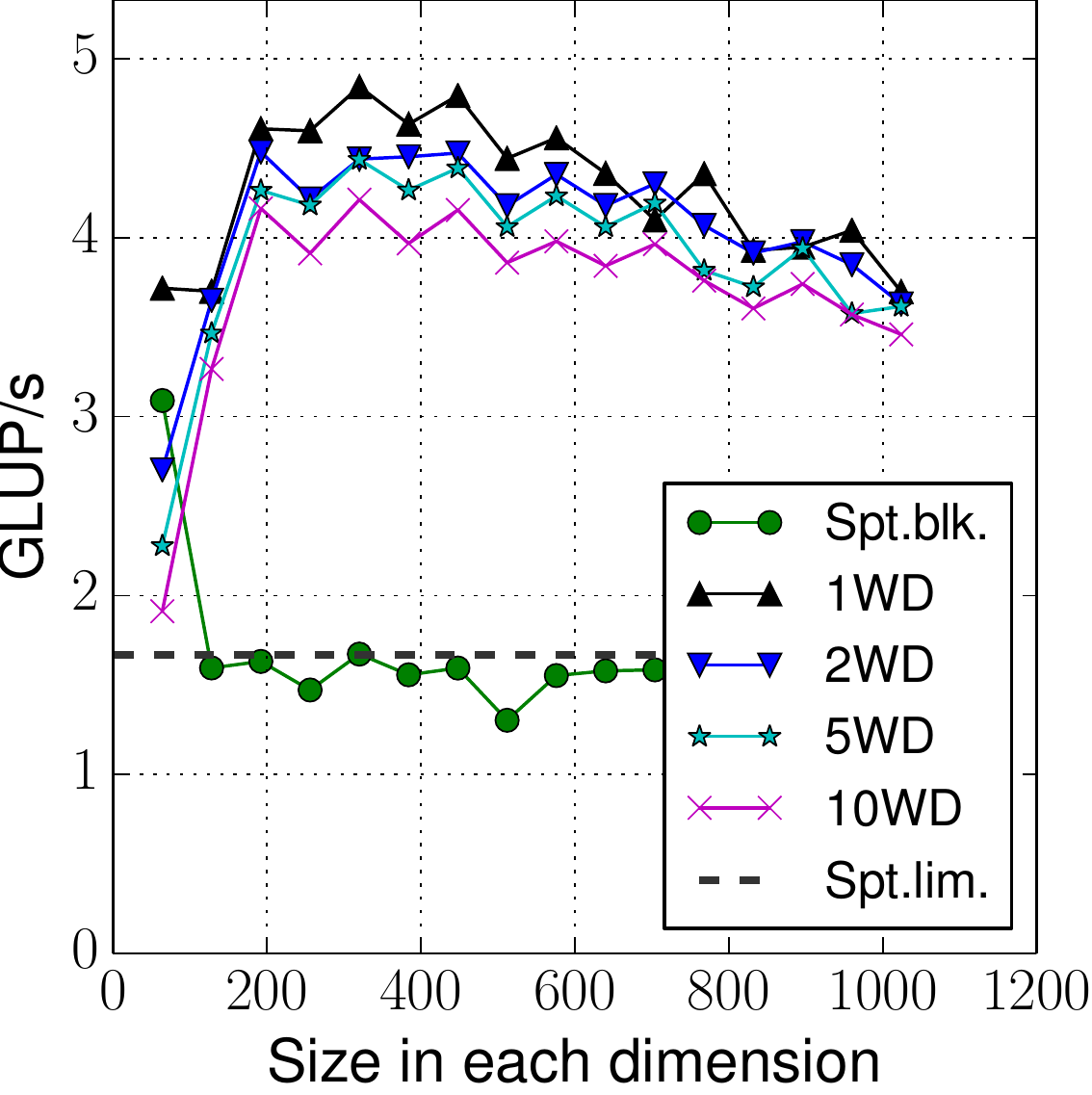}
        \label{fig:7_pt_const_strongscaling_perf}
    }
    \enskip
    \subfloat[7-point variable-coefficient stencil performance.]{
        \centering
        \includegraphics[width=3.9cm]{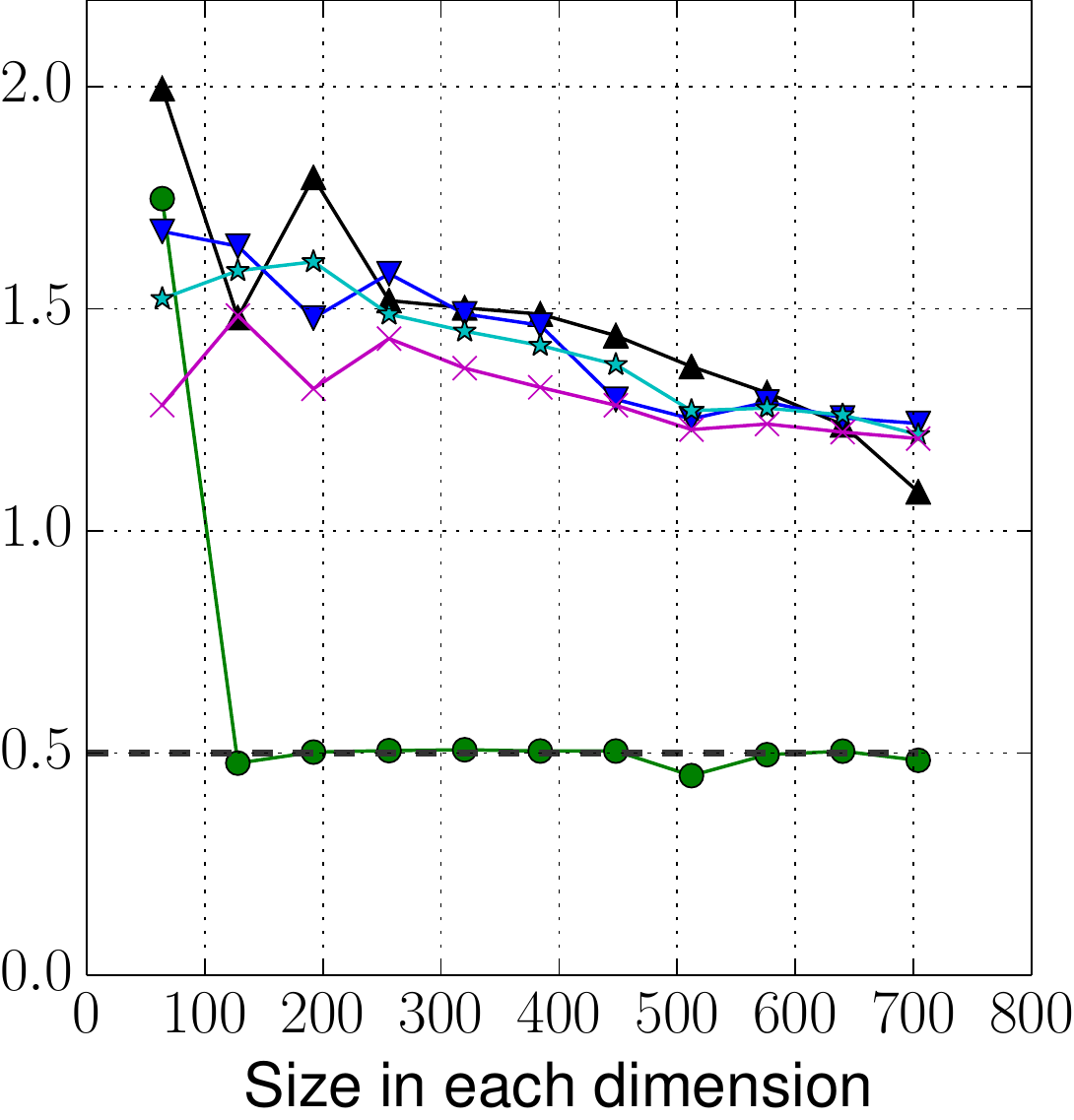}
        \label{fig:7_pt_var_strongscaling_perf}
    }
    \enskip
    \subfloat[25-point variable-coefficient stencil performance.]{
        \centering
        \includegraphics[width=3.9cm]{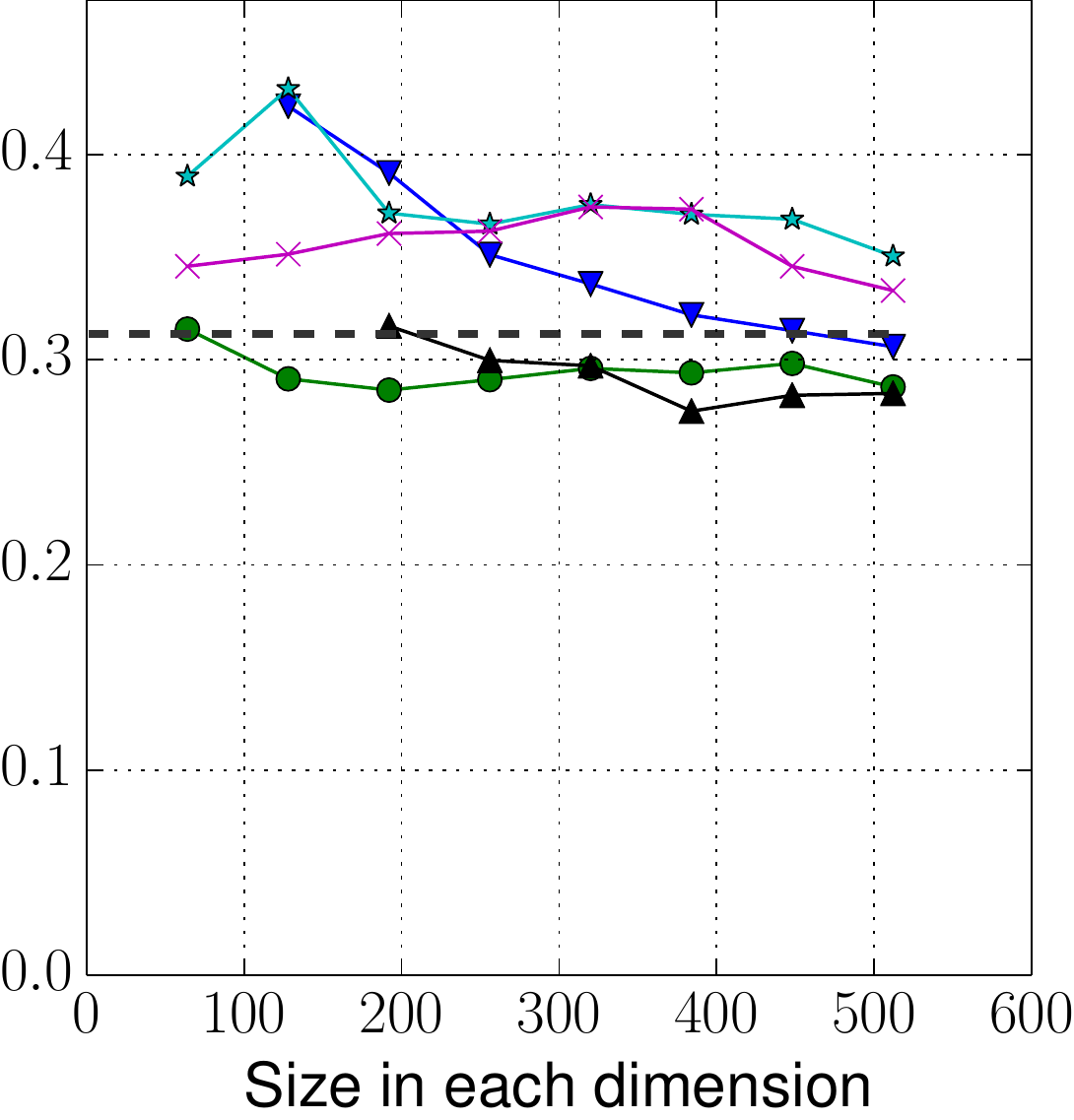}
        \label{fig:25_pt_var_strongscaling_perf}
    } 
     
    \subfloat[7-point constant-coefficient stencil measured memory bandwidth.]{
        \centering
        \includegraphics[width=4.15cm]{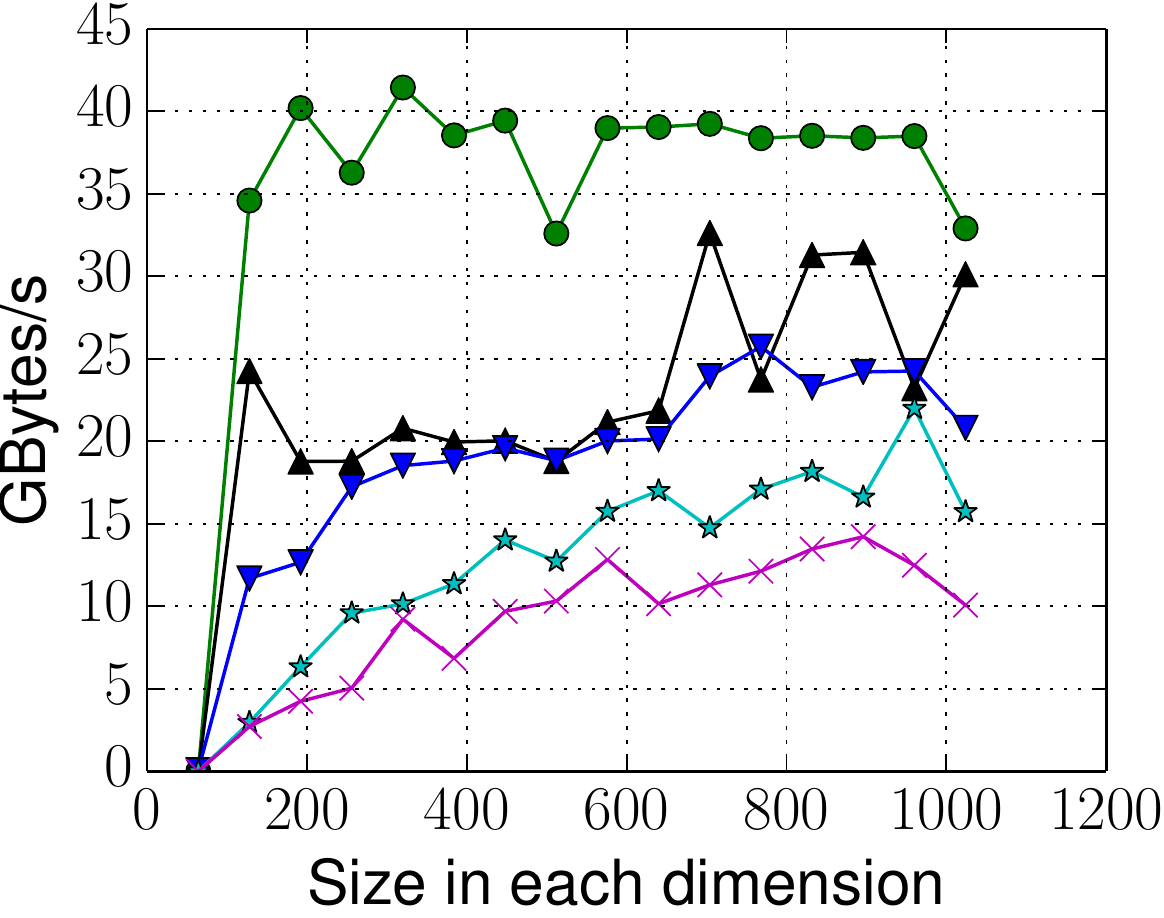}
        \label{fig:7_pt_const_strongscaling_bw}
    }    
    \enskip
    \subfloat[7-point variable-coefficient stencil measured memory bandwidth.]{
        \centering
        \includegraphics[width=3.8cm]{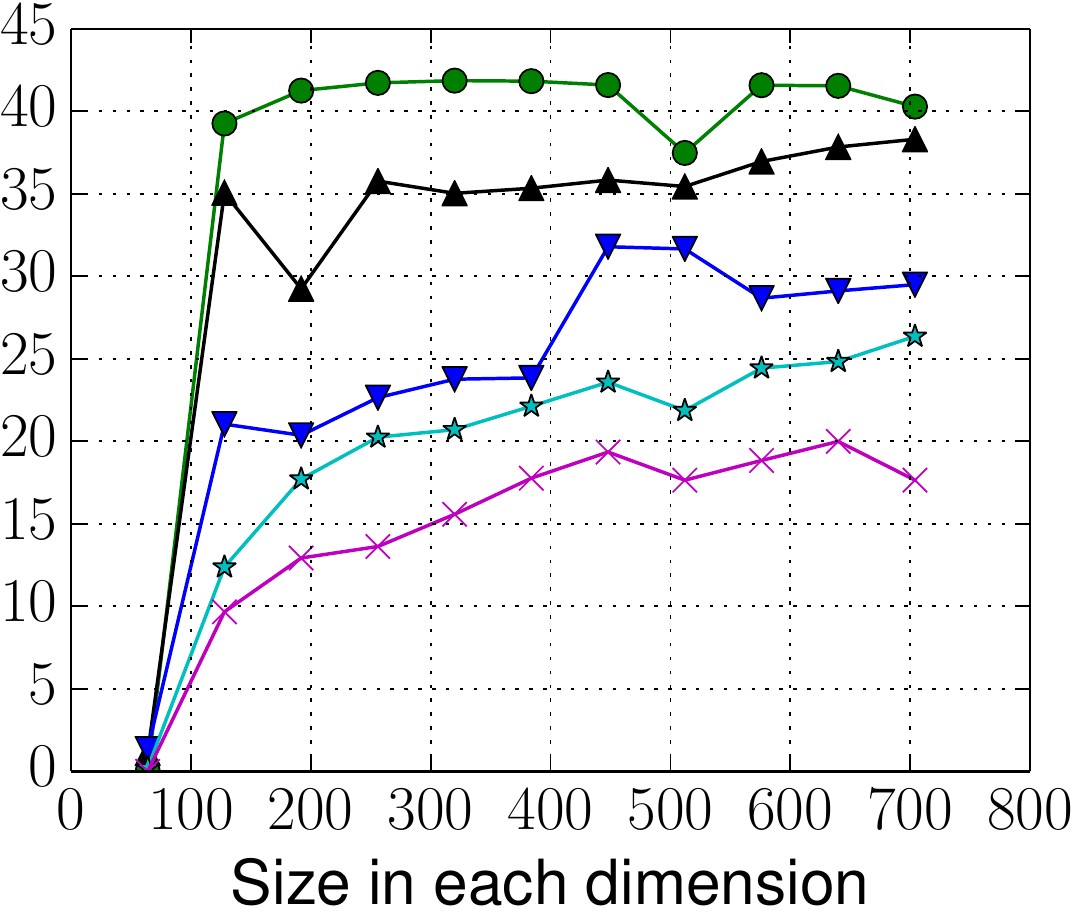}
        \label{fig:7_pt_var_strongscaling_bw}
    }
    \enskip
    \subfloat[25-point variable-coefficient stencil measured memory bandwidth.]{
        \centering
        \includegraphics[width=3.8cm]{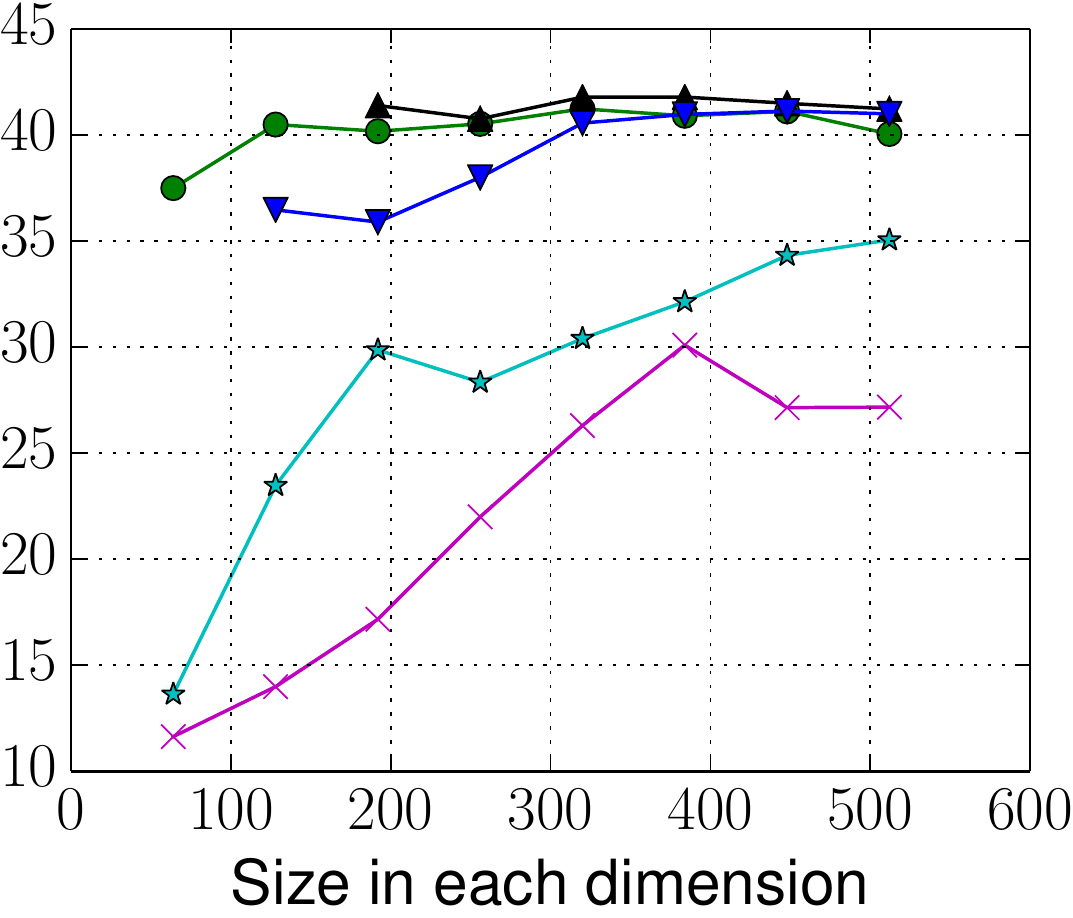}
        \label{fig:25_pt_var_strongscaling_bw}
    }    
    
    \caption{Performance (top row) and memory bandwidth (bottom row) 
      vs.\ grid size of three stencil types (columns) on a 10-core 
      Intel Ivy Bridge socket.  At 7-point constant-coefficient stencil, 
      all MWD algorithms 
      achieve between 2$\times$ and 3$\times$ speedup compared to the
      spatially blocked code, with 10WD using less than one third of
      the available memory bandwidth.}
    \label{fig:increasing_grid_size}
\end{figure*}

All results are shown in Figures~\ref{fig:7_pt_const_strongscaling_perf} and ~\ref{fig:7_pt_const_strongscaling_bw}.

1WD achieves the best performance at most of the grid sizes, as it decouples
from main memory and does not have the OpenMP overhead in the kernel,
which is described in the thread scaling results.  The MWD achieves
similar relative performance to 1WD as in the thread scaling results.
Also, larger thread groups consistently use less memory bandwidth at
all domain sizes as seen in
Figure~\ref{fig:7_pt_const_strongscaling_bw}.

The observed fluctuations in the performance are the results of the
variation in the selected diamond tile sizes at different grid sizes.
The requirement of having an integer number of diamond tiles in the
row of diamonds prevents the autotuner from selecting the optimal diamond
size.  Instead, it selects the nearest smaller diamond size that
allows integer number of diamonds in the given grid size.

In contrast to spatial blocking and 1WD, the other algorithms do not
achieve high performance for in-cache domain sizes.  This is mainly due to the 
OpenMP synchronization overhead, which becomes significant at small
problem sizes, and also probably
due to the high data sharing among the threads, which results in
invalidating many lines in the private caches.

\subsubsection{7-point stencil with variable coefficients}
All results are shown in Figures~\ref{fig:7_pt_var_strongscaling_perf} and \ref{fig:7_pt_var_strongscaling_bw}.

All the MWD approaches achieve similar performance at all grid sizes.
1WD outperforms the other variants at most of grid sizes up to a domain 
size of $512^3$, then 2WD takes the lead. 

\subsubsection{25-point stencil with variable coefficients}
All results are shown in Figures~\ref{fig:25_pt_var_strongscaling_perf} and ~\ref{fig:25_pt_var_strongscaling_bw}.

1WD does not have results at small domain sizes because there is no
sufficient concurrency for all the available threads, even with the
smallest diamond size ($D_w=16$).
2WD achieves the best performance up
to domain size $192^3$, where it starts hitting the memory bandwidth limit at larger grid sizes.

The performance-optimal diamond width selected by the auto-tuner varies 
with stencil type and grid size. Smaller grid sizes tend to get larger 
diamond tiles, as the leading dimension size is smaller. Stencils with 
variable coefficients have larger storage requirements per grid point, 
which limits the size of the performance-optimal diamond tile. At 
experiments with $N>100$, the median diamond width ranged between 16 and 32 
at the 7-point constant-coefficient, between 8 and 16 at the 7-point constant-coefficient, and between 16 and 32 at the 25-point variable-coefficient stencil.

\subsection{Distributed memory strong scaling performance}\label{sec:distmem}

\begin{figure*}[tbp]
    \centering
    \subfloat[Performance. Ideal scaling based on 2WD performance at single process.]{
        \centering
        \includegraphics[clip=true, trim= 0.0 -4.0 0.0 0.0, width=5.0cm]{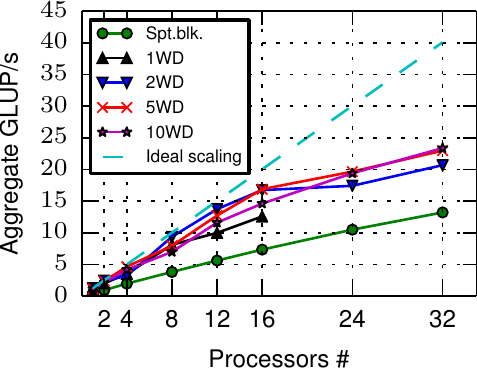}
        \label{fig:7_pt_var_dist_perf}
    }
    \enskip
    \subfloat[Time distribution. Error bars with standard deviation under 3\% are suppressed.]{
        \centering
        \includegraphics[width=7.3cm]{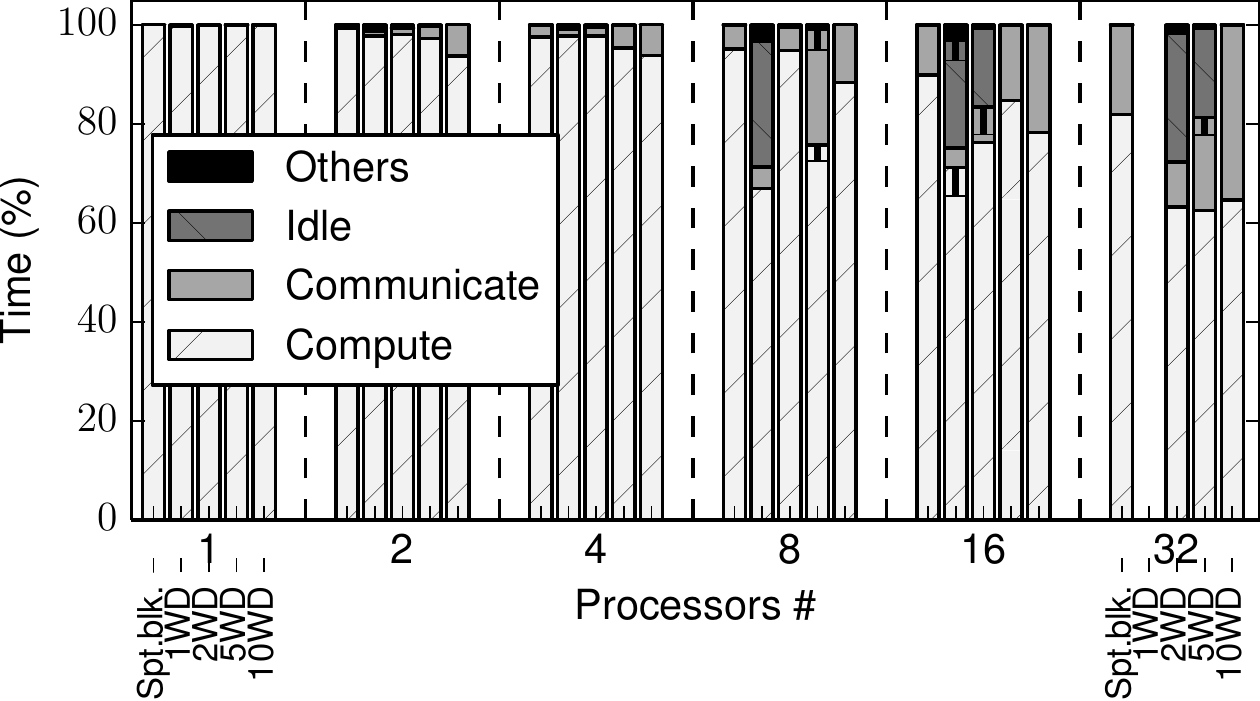}
        \label{fig:7_pt_var_dist_time_ratios}
    }
    \caption{Distributed memory strong scaling performance of the 7-point
      stencil with variable coefficients at a grid size of $768^3$.  A full
      10-core Intel Ivy Bridge socket was used per MPI process.}
    \label{fig:7_pt_var_dist}
\end{figure*}

Strong scaling performance experiments were performed for the 7-point
variable-coefficient and 25-point variable-coefficient stencils. We present
only the 7-point results here because they are not qualitatively
different from the 25-point case.
Domain decomposition across the $y$ axis is achieved through MPI message
passing, as described in Sect.~\ref{sec:approach:distmem}.  An Intel Ivy Bridge
socket is assigned to each MPI process, using ten OpenMP threads per
socket.
All results are shown in Figure~\ref{fig:7_pt_var_dist}.

1WD does not work beyond 16 processes because smaller subdomains in
the $y$ axis cannot provide sufficient concurrency to run all the
available threads (``concurrency condition'').  To run 1WD at 24 MPI
processes the minimum subdomain size would be 4 [min. diamond width] *
10 [threads/process] * 24 [processes] = 960 grid points along the $y$ axis.
Other MWD algorithms have less concurrency requirements in the diamond
tiling dimension, as another dimension of concurrency is introduced in
the wavefront blocking direction.  The concurrency condition can be
satisfied at 2WD, 5WD, and 10WD using 50\%, 20\%, and 10\% of the
minimum grid size that satisfies the concurrency condition at 1WD.
 
2WD achieves less performance compared to 5WD and 10WD at 24 and 32
processes due to the restrictions imposed by the concurrency condition,
which limits 2WD to small diamond tile sizes ($D_w=4$ at 24 and 32
processes, compared to $D_w=8$ at 16 processes).  On the other hand,
5WD and 10WD can use larger tile sizes while satisfying the
concurrency condition.  For example, 5WD and 10WD use $D_w=12$ and
$D_w=24$, respectively, at 32 processes.  The same concurrency
limitation causes the performance of 1WD to
drop at 12 and 16 processes, where $D_w=4$ compared to $D_w=8$ at 8
processes.
 
Timing routines are used in the code to profile the major parts.  As
shown in Figure~\ref{fig:7_pt_var_dist_time_ratios}, the run time is
mainly spent in performing stencil updates (``Compute''), communicating
the halo data across MPI processes (``Communicate''), and thread groups
idle time when the task queue is empty in the MWD implementation
(``Idle'').  Thread groups can have different time distribution as they
perform their tasks independently from each other.  Error bars are
used in Figure~\ref{fig:7_pt_var_dist_time_ratios} to present the
standard deviation of the thread groups' run time for each component.

When the concurrency limit is approached in the MWD implementation,
the idle time percentage increases, as can be seen in
Figure~\ref{fig:7_pt_var_dist_time_ratios}.  For example, the
subdomain of 1WD at 16 processes has 12 diamond tiles in the row
($N_y/(D_w\! \times \! P)=768/(4\! \times \! 16)=12$ diamond
tiles/row, where $P$ is the number of processes), which is very close
to the concurrency limit of the ten thread groups.  Handling
boundary tiles takes more time compared to interior tiles because of
the data exchange.  When the interior diamond
tiles of a row are updated before the boundary tiles, less concurrency
will be available.  This causes some thread groups to remain idle
when the subdomain size is near the concurrency limit.

As shown in Figure~\ref{fig:7_pt_var_dist_perf}, the MWD implementation
scales well up to 16 processes.  The large surface-to-volume ratio of
the subdomains at 24 and 32 processes results in large communication
overhead, as shown in Figure~\ref{fig:7_pt_var_dist_time_ratios}.
Performing domain decomposition at additional dimensions would allow
the code to have scalable performance at more processes, which is left
for the future work.

\section{Conclusion and future work}\label{sec:conc}

In this work we have combined the concepts of diamond tiling and
multi-core aware temporal wavefront blocking to construct
multi-threaded wavefront diamond blocking (MWD), a new temporal
blocking scheme for improving the performance of stencil computations
on contemporary multi-core CPUs. Using three different stencil types (short
range with constant coefficients, short range with variable
coefficients, and long range with variable coefficients) we have
demonstrated that our solution exerts considerably less pressure on
the memory interface for all stencils considered compared to
single-thread diamond tiling. It is also more flexible in terms of
assigning parallel resources to grid tiles since multiple threads work
concurrently on updates within a diamond tile using the shared
cache. Finally, the cache size requirements are also reduced. However,
depending on the diamond tile size, OpenMP and synchronization
overhead may be relevant, although we use a relaxed synchronization
scheme that avoids global barriers across tiles.
MPI-based distributed-memory parallelization of the MWD scheme is free
of global synchronization and lends itself to natural overlapping of
computation and communication.

Performance results on a modern 10-core Intel CPU show that MWD can
outperform single-thread diamond tiling when the code balance of the
purely spatially blocked code is large (such as for a 25-point long-range
stencil with variable coefficients). Future, more bandwidth-starved
architectures will benefit even more strongly from the reduced
pressure on the memory interface. We have also shown that the energy
consumption in the DRAM is considerably smaller with MWD, leading
to savings in energy to solution even if there is no performance gain.

These results open the possibility for future work in multiple
directions. First of all, our MPI implementation still suffers from idle
threads when boundary tiles in an MPI process are updated late. This
could be corrected by giving these tiles higher priority within the
subdomain. Furthermore, the one-dimensional domain decomposition
limits strong scalability and leads to a performance decay for large
problem sizes. It should be improved by introducing tiling
in the other dimensions. 

Some possible optimizations are marginal on mainstream multi-core CPUs
but may be decisive on future many-core CPUs like the Intel Xeon Phi,
where thread synchronization is costly and efficient SIMD
vectorization is absolutely required. Although some steps have already
been taken to reduce thread synchronization overhead, barriers are
still used inside the multi-core wavefront to ensure
correctness. These could be eliminated by a relaxed synchronization
scheme along the lines of Wittmann \textit{et al.}~\cite{wittmann10}. Finally,
alignment optimizations and data layout transformations could be imposed
to allow for fully aligned data accesses and thus enable more
efficient SIMD vectorization.

The implications of performance and memory transfer volume
for power dissipation and energy to solution have as yet not been
investigated thoroughly. As shown in Sect.~\ref{sec:7pt_const}, the MWD algorithms
show interesting power behavior because it is possible to end up in a
situation where energy to solution is minimal at a non-optimal
performance. Detailed energy models like the model introduced by Choi 
\textit{et al.}~\cite{Choi:2013:RME:2510661.2511392} may lead to valuable
insights concerning this issue.

\section*{Acknowledgments}

The Extreme Computing Research Center at KAUST supported T. Malas.
Part of this work was supported by the German DFG priority programme
1648 (SPPEXA) within the project Terra-Neo,
and by the Bavarian Competence Network for Scientific High Performance
Computing in Bavaria (KONWIHR) under project OMI4papps.

\bibliographystyle{siam}
\bibliography{references}

\end{document}